\documentclass[10pt,twocolumn,twoside]{IEEEtran}   
 \usepackage{microtype}
\usepackage{verbatim}
\usepackage[singlespacing]{setspace}
\usepackage{hyperref}
\usepackage{balance}
\usepackage{enumitem}
\setlist{leftmargin=3.5mm}
\usepackage{multirow}
\usepackage{amsthm}
\newcommand{\DM}{administrator\xspace}

\newcommand{\SPfull}{Secretary Problem\xspace}

\newcommand{\SSPfull}{Sequential Selection Problem\xspace}
\newcommand{\SSP}{SSP\xspace}
\newcommand{\WSSPfull}{Warm-starting Sequential Selection Problem\xspace}
\newcommand{\WSSP}{WSSP\xspace}
\newcommand{\DRA}   {DRA\xspace}

\newcommand{\RDRAshort}{Restricted DRA\xspace}
\newcommand{\RDRA}{RDRA\xspace}
\newcommand{\DRAfull}{Dynamic Resource Allocation\xspace}
\newcommand{\MSSPfull}{Multi-round Sequential Selection Process\xspace}
\newcommand{\MSSP}{MSSP\xspace}
\newcommand{\DP}{DP\xspace}
\newcommand{\DPs}{DPs\xspace}

\newcommand{\SDRAshort} {Sequential DRA\xspace}
\newcommand{\SDRA}   {SDRA\xspace}

\newcommand{\CCM}   {CCM\xspace}
\newcommand{\CCMstar}   {$\text{CCM}^*$\xspace}
\newcommand{\CCMfull}   {Cutoff-based Cost Minimization\xspace}
\newcommand{\MEAN}   {MEAN\xspace}
\newcommand{\RAND}   {RAND\xspace}
\newcommand{\MEDIAN}   {MEDIAN\xspace}

\newcommand{\sample}    		 {\sample\xspace}
\newcommand{\candidate}      		{candidate\xspace}
\newcommand{\candidates}    		 {candidates\xspace}
\renewcommand{\sample}    		 {sample\xspace}

\newcommand{\preselection}    		 {preselection\xspace}

\newcommand{\algo}    			 {\text{CCM}\xspace}

\renewcommand{\t}		{k}

\newcommand{\ninf}[1]	{\text{sum}(\Xbold{})}
\newcommand{\spaceInfection} 	{\{0,1\}}

\newcommand{\cost}    		 {\phi}

\newcommand{\policybold}   		 {\mathbf{\Pi}}
\newcommand{\access}    		 {\mathcal{C}}
\newcommand{\info}    		 {\mathcal{I}}
\newcommand{\nodes}    		 {\mathcal{V}}
\newcommand{\timeIndex}    		 {}

\newcommand{\cand}				 {}

\newcommand{\Xbold}[1]	               {\mathbf{X}_{#1}}
\newcommand{\val}			      {S} 
\newcommand{\Sbold}	               {\mathbf{\val}_{\cand}}

\newcommand{\X}[2]				{ X_{#1}}

\newcommand{\symb}[1]		   {#1^R}
\newcommand{\Spres}			{\symb{\val}}

\newcommand{\Cpres}	       		{\symb{C}}

\newcommand{\Sobold}		  	{ \mathbf{\Spres} }

\newcommand{\Abold}		        { \mathbf{R}_{\cand} } 
\newcommand{\A}[1]				{R_{#1}}

\newcommand{\Rbold}[1]    {\mathbf{R}_{#1}}

\newcommand{\C}[2]    {C_{#1,#2}}
\newcommand{\Co}[1]    {\Cpres_{#1}}
\newcommand{\Cbold}[1]    {\mathbf{C}_{#1}}
\newcommand{\Cobold}[1]    {\symb{\mathbf{C}}_{#1}}

\newcommand{\Rankset}[2] 			{\mathcal{P}_{#1}(#2)}

\newcommand{\Graph}    {\mathcal{G}}
\renewcommand{\ni}		{u} 
\newcommand{\mi}		{v} 
\newcommand{\diff}				{\Delta}
\newcommand{\numinfected}		{N^I}
\newcommand{\numinfectedoff}		{N^{I,\text{off}}}
\newcommand{\error}		{e}
\newcommand{\adjbold}		{\mathbf{A}}
\newcommand{\adj}		     {A}
\newcommand{\cone}			{c_1}
\newcommand{\ctwo}			{c_2}

\newcommand{\emphd}[1] 	{#1}
\usepackage{graphics}
\usepackage{trimclip}
\usepackage{algorithm}
\usepackage{algorithmicx,algpseudocode}

\newtheorem{definition}{Definition}
\newtheorem{remark}{Remark}
\newtheorem{assumption}{Assumption}
\usepackage{xr}
\usepackage{hyperref}
\usepackage{url}
\usepackage{amsmath}
\usepackage{amssymb}
\usepackage{amsfonts}
\usepackage{dsfont}
\usepackage{upgreek}
\usepackage{bbm}			
\usepackage{graphicx}
\usepackage{epstopdf}
\graphicspath{ {./figures/} }
\usepackage{tikz}
\usepackage{float}
\usepackage{booktabs} 
\usepackage{subfigure}
\usepackage{xspace}
\usepackage[scaled=.7]{beramono}
\makeatletter
\g@addto@macro\bfseries{\boldmath}
\makeatother
\newcounter{phase}[algorithm]
\newlength{\phaserulewidth}
\newcommand{\setphaserulewidth}{\setlength{\phaserulewidth}}

\makeatother
\setphaserulewidth{.3pt}
\newcommand{\Sec}[1]		{Sec.\,\ref{#1}}
\newcommand{\Fig}[1]		{Fig.\,\ref{#1}}

\newcommand{\Figs}[1]   {Figs.\,\ref{#1}}
\newcommand{\Eq}[1]			{Eq.\,\ref{#1}}
\newcommand{\Tab}[1]		{Tab.\,\ref{#1}}
\newcommand{\Alg}[1]		{Alg.\,\ref{#1}}

\newcommand{\Remark}[1]{Remark\,\ref{#1}}

\newcommand{\Definition}[1]{Definition~\ref{#1}}

\newcommand{\Assumption}[1]{Assumption~\ref{#1}}

\newcommand{\vs}   			{vs.\@\xspace}
\newcommand{\ie}   			{i.e.\@\xspace}
\newcommand{\eg}   			{e.g.\@\xspace}
\newcommand{\etc}   		{etc.\xspace}
\newcommand{\wrt}   		{w.r.t.\@\xspace}

\newcommand{\st}   			{\mbox{s.t.}\xspace}
\newcommand{\Erdos}   	{Erd\"os-R\'enyi\xspace}
\newcommand{\one}       {\mathbf{1}}     
\newcommand{\zero}      {\mathbf{0}}
 
\newcommand{\ind}       {\mathds{1}}
\newcommand{\Ind}[1]    {\mathds{1}{\{#1\}}}

\newcommand{\Exp}[1]    {\mathbb{E}[#1]}
\newcommand{\Prob}      {\mathbb{P}}

\newcommand{\real}     {\mathbb{R}}

\newcommand{\natnostar} {\mathbb{N}}
\newcommand{\nat}     {\mathbb{N}^*}
\newcommand{\pdf}				{p.d.f.\xspace}

\newcommand{\mydots} 	{...}
\newcommand{\algComment}[1] 	{\hfill/\!/\,{#1}}
 
\newcommand{\inlinetitle}[2]  {\vspace{4pt}\noindent\textbf{\emph{#1}{#2}}}
\newcommand{\inlinetitleemph}[2]  {\vspace{4pt}\noindent{\emph{#1}{#2}}}

\renewcommand*{\top}{{\mkern-1.5mu\mathsf{T}}}

\DeclareMathOperator*{\argmax}{\arg\!\max}

\newcommand{\obullet}				{\square}
\newcounter{marginNoteCounter}

\begin{document}
\title{\LARGE \bf Dynamic Epidemic Control via Sequential Resource Allocation}
\date{}
\author{Mathilde Fekom \quad Nicolas Vayatis \quad  Argyris Kalogeratos
\thanks{\!\!\!\!\!\!$\obullet~$Contact emails:
{\tt\footnotesize  \{fekom,\,vayatis,\,kalogeratos\}@cmla.ens-cachan.fr}.}%
\thanks{\!\!\!\!\!\!$\obullet~$Telephone number:
{\tt\footnotesize + 33 1 47 40 20 00.}}%
\thanks{\!\!\!\!\!\!$\obullet~$Mailing address:
{\tt\footnotesize  4 avenue des Sciences
91190 Gif-sur-Yvette, France.}}%
\thanks{\!\!\!\!\!\!$\obullet$~Part of this work was funded by the French Railway Company, SNCF, and the IdAML Chair hosted at ENS Paris-Saclay.
}
\\
\vspace{2mm}
\emph{Universit\'{e} Paris-Saclay, ENS Paris-Saclay, CNRS, Centre Borelli, 91190 Gif-sur-Yvette, France}\\
}

\maketitle
\begin{abstract}
In the \emph{Dynamic Resource Allocation} (DRA) problem, an \DM has to allocate a limited amount of resources to the nodes of a network in order to reduce a diffusion process (\DP) (\eg an epidemic).
In this paper we propose a multi-round dynamic control framework, which we realize through two derived models: the \emph{Restricted} and the \emph{Sequential \DRA} (\RDRA, \SDRA), that allows for restricted information and access to the entire network, contrary to standard full-information and full-access \DRA models. At each intervention round, the \DM has only access --simultaneous for the former, sequential for the latter--  to a fraction of the network nodes. This sequential aspect in the decision process offers a completely new perspective to the dynamic \DP control, making this work the first to cast the dynamic control problem as a series of sequential selection problems. Through in-depth SIS epidemic simulations we compare the performance of our multi-round approach with other resource allocation strategies and several sequential selection algorithms on both generated, and real-data networks. The results provide evidence about the efficiency and applicability of the proposed framework for real-life problems.
\end{abstract}

%=====================================================
\section{Introduction}\label{sec:intro}
%=====================================================

Compartmental models have gained particular attention in recent years due to their simple analytic formulations that can model modern problems related to information diffusion and social epidemics, \eg rumor spreading \cite{Jin2013, rumorSpreading2018} and other social contagions \cite{SISa_obesity}. Controlling efficiently undesired diffusion processes (\DPs) is crucial for public security and health, as was dramatically illustrated also by the Covid-19 crisis \cite{Ezekiel20}. Yet, it is a difficult problem that in fact gets instantly much more complicated the moment one starts including more realistic constraints or objectives. 
A source of limitations is the theoretical \emph{interaction model} to consider along with its network-wise abstraction (\eg macro- vs microscopic modeling), which may be over-simplistic for the analyzed phenomenon. 
Another source of shortcomings is the level of required information regarding the \emph{system state}, such as the infection state of nodes or the network connectivity. 
Finally, limitations come from the way a control model assumes it can intervene to the \DP, \eg in a fixed or dynamic fashion to the evolution of the process.

Dynamic models for allocating medical resources are subject to wide investigation \cite{Chen18, Liu13}, for which \cite{Scaman15} gives a convenient formalism with the introduction of the \emph{Dynamic Resource Allocation} (DRA), a model for network control, originally developed for SIS-like processes \cite{Mieghem09} (a node is either infected, or healthy without permanent immunity), that distributes a limited budget of available treatment resources on infected nodes in order to speed-up their recovery. The \emph{score-based} DRA formulation introduces an elegant way for assessing, through a score value, the criticality of each node individually for the containment of the DP \cite{Scaman15,Scaman16}. 
Then, the \DM only has to ensure that at each moment the resources will be spent on the infected nodes with the highest scores. Among the proposed options \cite{Tong12, Schneider11}, a simple yet efficient local score is the \emph{Largest Reduction in Infectious Edges} (LRIE) \cite{Scaman15}, which depends on the infection state of the neighbors, hence it needs to be updated regularly during the process.
A second option, called \emph{priority planning} \cite{Scaman16}, computes offline a \emph{priority-order} of the network nodes so that to have minimal \emph{max-cut}. This order specifies a fixed global score for all nodes, which can then be used to perform DRA considering only the infected nodes each time. 
In \cite{Lorch18}, scores are continuous (called \emph{control signals}) and are derived for each node with the purpose of minimizing a loss function representing  trade-off between the cost of a treatment and the cost incurred by an infected node.

\begin{figure}
\centering
\hspace{-5mm}
\includegraphics[width=0.59\linewidth, viewport=10 2 380 130,clip]{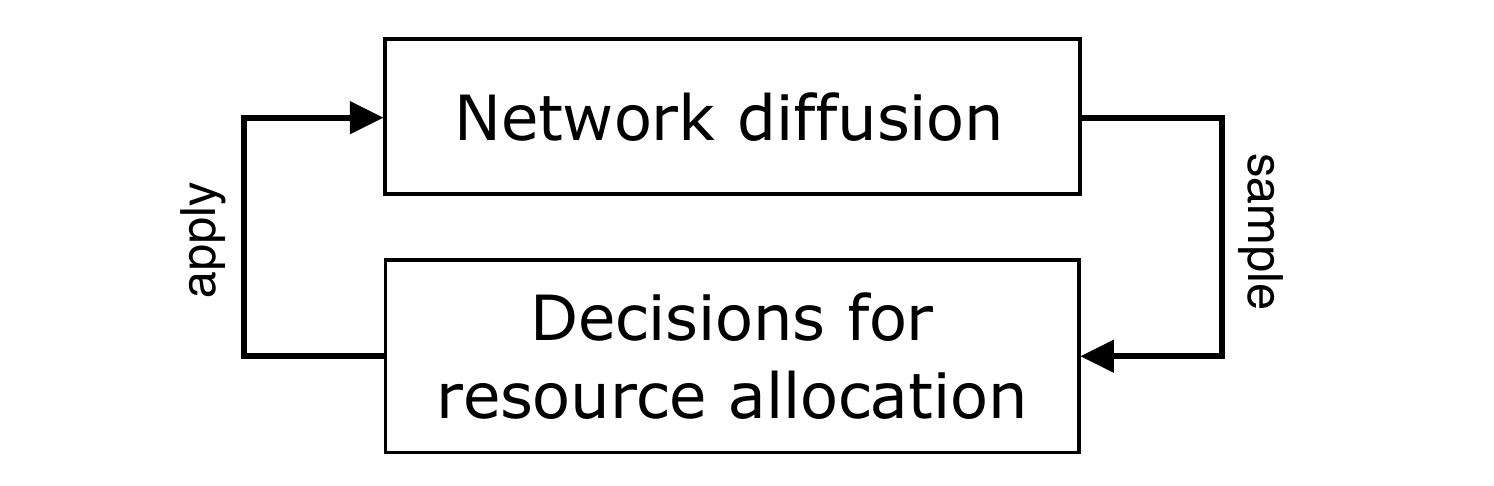}
\vspace{-0.5em}
\caption{Scheme of the network diffusion control problem.}
\vspace{-1.5em}
\end{figure}
The motivation of our work is to bring the score-based DRA modeling closer to reality. In real-life scenarios, authorities have access to limited information regarding the network state, and can reach a limited part of the population to apply control actions (\eg deliver treatments). Even more importantly, the decision making process is essentially a sequence of time-sensitive decisions over choices that appear and remain available to the \DM only for short time, also with little or no margin of revocation. 
An intuitive paradigm to consider is how a healthcare unit works: patients arrive one-by-one seeking for care, and online decisions try to assign the limited available resources (\eg medical experts, beds, treatments, intensive care units) to the most severe medical cases \cite{Bekker17, Kabene06, Gnanlet09}. While dealing with the Covid-19 crisis, medical experts in many countries were in fact `scoring' patients to prioritize hospitalization. Above all, the unprecedented stress induced by the pandemic to the healthcare systems did reveal several of its weaknesses, while also highlighted the lack of efficient support for sequential decision making. The subject of this work aims to fill methodological gaps exactly of this kind. 
We establish a link between the \DRA problem and the sequential decision making literature, offering a completely new perceptive to dynamic \DP control. Among the existing \emph{\SSPfull}\emph{s} (\SSP) that have been widely studied, the most well-known is the \SPfull \cite{Ferguson89}. Our aim, however, is to propose a concordant match to the discussed control setting.

Concerning the technical contribution, we first present the \emph{\RDRAshort} (\RDRA) model, in which each time the \DM can decide the reallocation of the resources only among a random sample of currently reachable nodes, which is treated as a batch. 
We next propose the special case of \emph{\SDRAshort} (\SDRA) where the latter sample of nodes is provided with a random \emph{arrival order}, forcing the \DM to decide for the resource reallocation sequentially, according to the characteristics of the incoming nodes. 
The major achievement of our modeling is that it manages to create a new playground
\footnote{A preliminary short version of this research was presented in \cite{Fekom19bis}.}
 where \SSP algorithms can be incorporated to the \DP control and make control strategies more applicable in real conditions. The implementation of existing online algorithms, such as the \emph{hiring-above-the-mean} \cite{Broder09} or even the more effective \emph{\CCMfull} \cite{Fekom19}, leads to \SDRA strategies that manage to reduce a \DP in a comparable fashion to the unrestricted \DRA.
%=====================================================
\section{\DRAfull strategies}\label{sec:setting}
%=====================================================
%--------------------------------------------------------------------------------------------------
\subsection{Network diffusion process} %--------------------------------------------------------------------------------------------------
The interactions among a population of $N$ individuals are modeled by a fixed network represented by a graph $\Graph(\nodes,\mathcal{E})$ of $|\nodes| = N$ nodes and $|\mathcal{E}| = E$ edges. The graph structure is arbitrary; the reader might picture it as being a directed or non-directed, weighted or unweighted graph,  etc. Each entry $\adj_{ij}$ of the graph's adjacency matrix $\adjbold \in \real^{N\times N}$, expresses the possible influence of node $i$ on node $j$, \st $\adj_{ij} \neq 0$ if node $i$ is linked with an edge to node $j$ and may have an impact on him, and vice-versa.

The graph hosts an \emph{agent-based diffusion process} (\DP) that spreads from one node to another. Conceptually, the sequential epidemic control framework that we present, can apply to arbitrary \DPs (even non-compartmental diffusion models), provided that there is a means to assess the criticality of the graph elements (\eg nodes) for reducing the epidemic process. 
To simplify the presentation, we use a setting for which such criticality assessment has been made possible in the literature. We suppose that the \DP in place is a continuous-time Markov process \cite{Mieghem09}, so that at each time instance $t\in \real_+$ there can be at most one event of node state change in the network. More specifically, we consider an SIS-like recurrent epidemic, where nodes are either healthy (`S': susceptible to infection), or infected (`I'). The infection spreads from any infected node to its reachable healthy neighbors. Nodes are equipped with self-recovery without ever achieving permanent immunity. 
The infection state of the network is denoted by $\mathbf{X}_{t}=(X_{1,t},\mydots,X_{N,t})^\top \in \spaceInfection^N$, \st $X_{i,t}=1$ if node $i \in \nodes$ is infected and $0$ otherwise. In the rest of the paper, $\bar{X}_{i,t} = 1 - X_{i,t}$, and $\numinfected_t= \sum_{i=1}^N X_{i,t}$ stands for the number of infected nodes in the network at time $t$. 

%--------------------------------------------------------------------------------------------------
\subsection{Resource allocation for controlling a \DP} \label{sec:SIS-DRA}
%--------------------------------------------------------------------------------------------------
An \emph{\DM} has the mission to reduce the DP by managing a fixed budget of $b \in \nat$ resources that help the receiving nodes leaning towards the healthy state. 
The resources are regarded as being reusable, non-storable through time, and non-cumulable at nodes (\ie at most one on each node). Resources might serve as treatments, doctors, nurses, beds, \etc in an emergency service, in which allocation decisions have to be made on-the-go.
At each time instance $t\in\real_+$, the \emph{\DRAfull} (DRA) \cite{Scaman15, Scaman16} dynamically determines the resource allocation vector $\Rbold{t}= (R_{1,t},\mydots,R_{N,t})^\top \in \{0,1\}^N$ where $R_{i,t} =1$ if a treatment is allocated to node $i$ at time $t$ and $0$ otherwise; subject to $\sum_i R_{i,t} = b$.
The node state transitions of the \emph{homogeneous SIS model under control} are given by:
\begin{align}
\begin{split}
\text{(healthy to infected)}\hspace{3mm} X_{i,t}&: 0 \rightarrow 1 \  \text{at rate} \ \beta \textstyle \sum_j \adj_{ij} X_{j,t}; \\
\text{(infected to healthy)}\hspace{3mm} X_{i,t}&: 1 \rightarrow 0 \ \text{at rate} \ \delta + \rho R_{i,t},
\end{split}
\label{eq:rates}
\end{align}
where $\beta$ is the contribution of any edge to the infection rate from an infected towards a healthy node, $\delta$ is the self-recovery rate of each infected node, and $\rho$ is the contribution of a received treatment (if $R_{i,t} = 1$) to the node recovery rate. %
If we let $\lambda_i$ be the transition rate for node $i$ in \Eq{eq:rates}, then the mechanism generating \DP events (infection or recovery) is a Poisson process and the probability distribution of the time intervals between events is $p(t,\lambda) = \lambda e^{-\lambda t}, \,\, \forall t \in \real_+$, where $\lambda = \sum_{i=1}^N \lambda_i$ is the sum of all node transition rates. Then, each node $i\in \nodes$ can be the transitioning node with probability equal to $\lambda_i/\lambda$. 
Using the formalism of \cite{Allen08}, we write as $p_{\mi \ni}(\Delta t) := \Prob(\numinfected_{t+\Delta t}=\mi \,|\, \numinfected_t = \ni)$ the probability of going from $\ni$ to $\mi$ number of infected nodes in a time interval $\Delta t$, and we get:
\vspace{-1mm}
\begin{align}
\!\!\!\!p_{\mi \ni}(\Delta t ) = 
\begin{cases}
  b(\ni) \Delta t &  \mi = \ni +1\\
 d(\ni) \Delta t &  \mi = \ni - 1\\
 1 - \left(b(\ni)+d(\ni)\right)\Delta t & \mi = \ni\\
 0 & \mi \notin \{\ni+1,\ni,\ni-1\},\!\!\!\!\!\!
\end{cases}
\label{eq:transition} 
\end{align}
\begin{equation}
\hspace{-6.2em}\text{where\ } b(\ni) := b(\numinfected_t = \ni) = \textstyle \beta \sum_{ij} \adj_{ij} \bar{X}_{i,t}X_{j,t},
\end{equation}
\vspace{-1.5em}
\begin{equation}
\hspace{-3.4em}d(\ni) := d(\numinfected_t = \ni) = \textstyle \sum_i(\delta + \rho R_{i,t})X_{i,t}.
\end{equation}
For continuous-time, $\Delta t \rightarrow 0$, and by using the Markov property $\Prob(\numinfected_{t+\Delta t} \,\,|\,\, \numinfected_0, ...,\numinfected_t) = \Prob(\numinfected_{t+\Delta } | \numinfected_t)$, the forward Kolmogorov differential equations are found for the probability of having $n$ infected nodes at time $t$, denoted by $p_\ni(t)$:
\begin{equation}
\frac{d p_{\ni}(t)}{d t} = p_{\ni-1}b(\ni -1) + p_{\ni+1} d(\ni+1)- p_\ni(b(\ni)+d(\ni)),
\label{eq:kolmo}
\end{equation}
for  $n= 1,2,...,N$, and $dp_0/ dt=p_1d(1)$.
This standard equation enables us to get the evolution of the stochastic variables of the problem. In particular, by multiplying by $n$ and summing over $n$ we find the equation of the evolution of the epidemic: 
\begin{equation}
\frac{d\Exp{\numinfected_t}}{d t} = \beta \,\,\Exp{ \mathbf{{X}}^\top \adjbold \mathbf{\bar{X}} } - \delta \Exp{\numinfected_t} - \rho \Exp{ \mathbf{X}^\top \mathbf{R}},
\end{equation}
where $\Exp{ \mathbf{X}^\top \mathbf{R}} = \min(b, \numinfected_t)$.

%--------------------------------------------------------------------------------------------------
\subsection{Scoring function}
%--------------------------------------------------------------------------------------------------
The \emph{score-based DRA} assumes that there exists a scoring function $s: \nodes \rightarrow \real$ that returns a score $S_{i,t} \in \real$ for each node $i$ at time $t$, according to the mission, and the nodes with the highest scores are those to receive the resources. This class of strategies depends on the size of the available budget of resources and the efficiency of each of them, as well on the ability of the scoring function in assessing correctly the criticality of nodes. 

The evaluation of the performance of a score-based strategy at time horizon $T\in \real_+$ is the expected area under the curve (AUC) of the percentage of infected nodes \wrt time:   
\begin{align}
{A_N(T)} &= \int_0^{T}\frac{\Exp{\numinfected_t}}{N} dt\,\,\, \in \real_+, \label{eq:auc}
\end{align}
By making this choice we acknowledge that the $\Exp{A_N(T)}$ is more characteristic for the effect of a strategy than, for example, the expected extinction time, $\Exp{t_{\text{exct.}} } $.
%
%================================
\section{Sequential \DRAfull}
%================================
\begin{figure}[t]
\centering
\hspace{-2.8em}
\includegraphics[width=0.99\linewidth, viewport=70 50 731 739,clip]{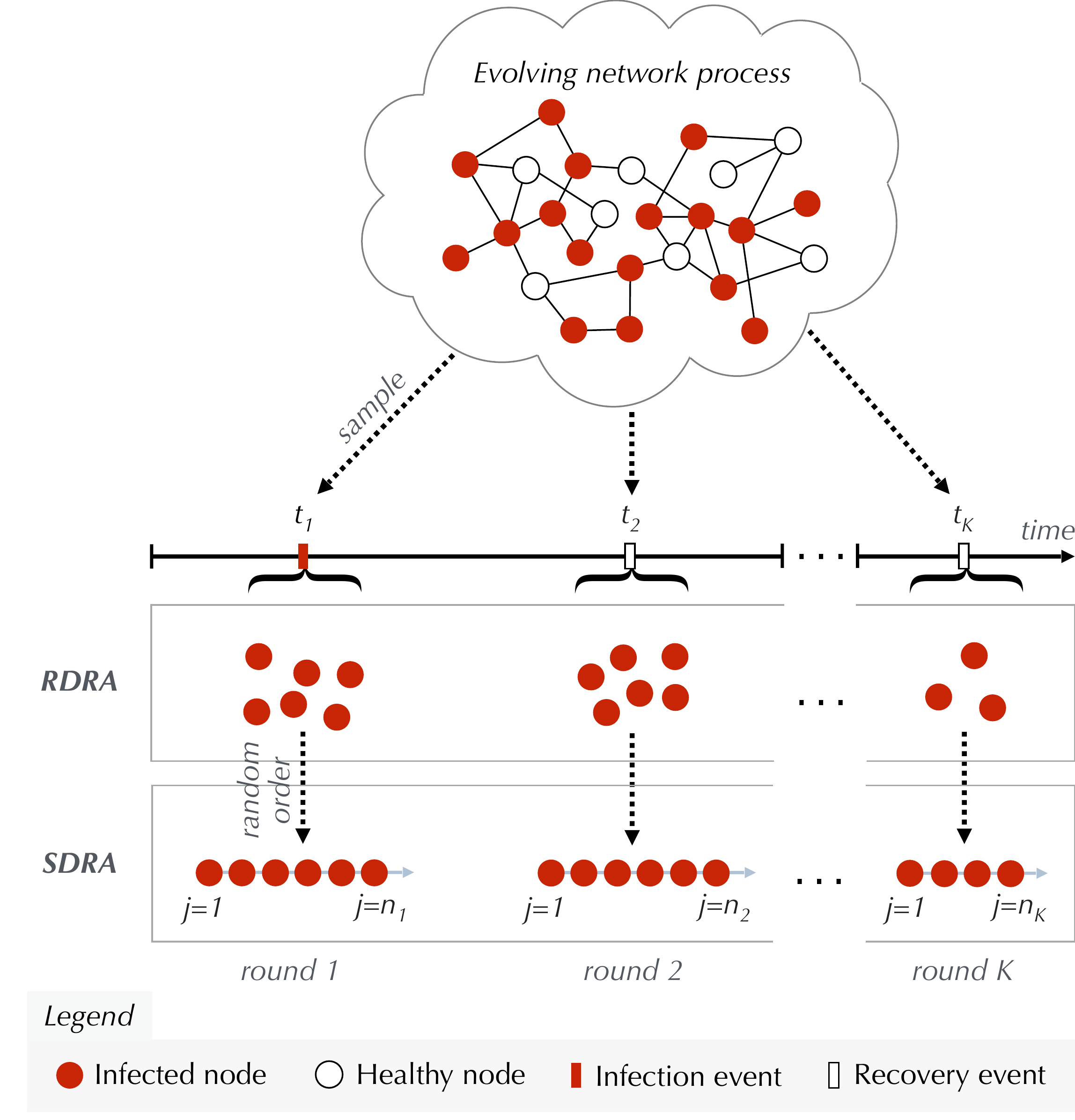} 
\vspace{-0.5em}
\caption{The sequential evaluation of candidates in the \SDRA model.}
\label{fig:time_scale}
\end{figure}

The standard \DRA strategies are build on a strong assumption whereby the \DM has always full information about the process and full access to the network, which is apparently far from being realistic in most practical cases. To reduce this distance, we introduce two models in the following sections. Their generality stems from the fact that they assume the scoring function $s$ to be a `black-box' (hence out of our main research focus here) that is appropriate for the studied network process. In this context, this means that it is efficient in determining the criticality of an infected node when asked. It then remains to the algorithmic part of a strategy to take the decisions of resource reallocations to nodes that would be as much as possible valued by the function $s$.

%==============================
\subsection{Restricted \DRA}\label{sec:RDRA}
%==============================
%
In the \emph{\RDRAshort} (\RDRA) model, only a fraction of nodes are reachable at each moment. Let $\info$ be the set of nodes for which we have information and $\access$ the set of accessible nodes, with $\info, \access \subseteq \mathcal{V}$.
We put forward two reasonable assumptions: 
A1)~the accessible nodes are always included in the set of nodes for which we have information, \ie $\access \subseteq \info$, and A2)~at any time $t \in \real_+$, the treated nodes, $\Cobold{t} = \{i \in \nodes : R_{i,t} = 1\}$ are accessible.
The first notion to define for specifying the \RDRA model is that of a \emph{round}. %
\begin{definition}\emphd{Round} $k$ is a discrete event of reviewing and revising the allocation of resources on the network. The series of rounds is defined by the sequence of time instances $(t_k)\in\real_+^K$, where $t_{0} = 0$, $t_K = T$, and $K$ is the total number of rounds. 
\label{def:round}
\end{definition}

\begin{remark}\label{rmk:round-triggering}
A round-triggering process is considered to invoke the revision of the resource allocation. In this paper we restrict ourselves to a passive process, \ie the $(t_k)'s$ are not decided or known to the \DM in advance. However, an active process can be an interesting addition to the \RDRA strategy.
\end{remark}

Generally, two subsequent rounds can be arbitrarily distant in time. Here, more specifically, a new round is triggered whenever there is a change in the infection state of the network, and thus, at most one node can recover or get infected between two subsequent rounds. 
Formally, this is described by the following condition on the allocation events:
\begin{equation}
t_{k} = t_{k-1}+ \text{min}(\delta t \mid ||\Xbold{t+\delta t} - \Xbold{t} ||=1),\, \ \forall k \leq K.
\end{equation}
Since rounds is essentially a measure of time, each variable can be defined round-wise, for instance we write $\Xbold{k}$ for the infection state at round $k$, \ie at the time instance $t_k$.

\begin{definition}The \emphd{Restricted \DRA} (\RDRA) strategy $\policybold_{k}(\info,\access)$ is a DRA strategy parametrized by the number of resources $b \in \nat$ and a scoring function $s:\nodes\rightarrow \real$. At any moment during a round $k$, the \DM has information about the nodes of the set $\info$ and simultaneous access to the nodes of the set $\access$, in addition to those currently treated, $\Cobold{k}$, more specifically: $\Cobold{k} \subseteq \access$. The strategy outputs a resource allocation vector, \ie $\policybold_{k} = \Rbold{k}
,  \, \forall k \le K$.
\label{def:rdra}
\end{definition}
\noindent
We consider as the default RDRA setting when $\info = \access = \Cbold{t}$, where $\Cbold{t}$ is called \emph{sample} and is defined below. Choosing to define $\access$ or $\info$ otherwise, results in special \RDRA cases, {while choosing $\info= \access = \nodes$ degenerates to standard DRA.}

\begin{definition}{\emphd{Node sample} $\Cbold{k} $} is the set of accessible infected nodes at time $t_k$, $\Cbold{k} =(\C{1}{k},\mydots,\C{n_k}{k}) \subset \nodes$. Its size $n_k=f(\Xbold{k}) \in \nat$ is given by $f:\spaceInfection^N \rightarrow \nat$, a function of the infection state. 
The probability of observing a sample $c \subset \nodes$, given its size $n\in \nat$ and the network state $x\in \spaceInfection^N$, is $\Lambda_{k}(c;n,x)=\Prob(\Cbold{k}=c \mid |\Cbold{k}| = n, \Xbold{k} = x)$. In short we write $\Cbold{k} \sim \Lambda_{k}(n,\Xbold{k})$. 
\label{def:sample}
\end{definition}

Typically, the number of accessible nodes is proportional to the number of infected nodes in the graph, \ie $n_k = \lfloor \alpha \sum_iX_{i,k}\rfloor$, $\alpha \in [0,1]$. 
\noindent Take as examples these two simple sampling functions, whereas more complicated ones can be considered: 
\begin{itemize}
\item $\Prob(i \in \Cbold{k}) =  \frac{n_k}{N}\Ind{X_{i,k}=1}$, \ie the \vspace{1mm}  sampling is uniform among the population of infected nodes;
\item $\Prob(i \in \Cbold{k}) = frac{e^{S_{i,k}} }{ \sum_i e^{S_{i,k}}}\Ind{X_{i,k}=1}$, \ie \vspace{1mm} nodes with high scores have a highest probability to be sampled.
\end{itemize}
The \RDRA's assumption for simultaneous access to all the nodes of a sample in a round remains far from being realistic. We next refine the access constraints and present a model processes sequentially the sample.
%=================================
\subsection{Sequential \DRA}
%================================= 
The \emph{\SDRAshort} (\SDRA) model does not reassign altogether the $b$ treatments as \DRA and \RDRA strategies do. Here, the round is further divided into $n_\t$ time intervals and the reallocation is performed sequentially at the time-scale of the round duration. In each time interval one candidate of the sample is examined for getting a treatment. Let the discrete index $j \in \{1,\mydots,n_k\}$ characterize the sequential arrival order of candidates, \eg $j=1$ and $j=n_k$ are respectively the first and last candidates of round $k$.
Since the \DM gains access sequentially to candidates, the problem variables can depend on the index $j \in \{1,\mydots,n_k\}$.

\begin{definition}{The \emph{\SDRAshort} (\SDRA) $\policybold_{k}(\info, \access)$} is the \RDRA strategy defined by the sequence $\policybold_{k}(\info, \access) = (\policybold_{k,1}(\info_1, \access_1), ..., \policybold_{k,n_k}(\info_{n_k}, \access_{n_k}))  \in \{0,1\}^{n_k}$ where $\info_j=(C_{1}, ...,  C_{j})$ and $\access_j = C_{j}, \,\, \forall j \leq n_k$, providing a uniformly random arrival order of the nodes of the sample.
\label{def:sdra}
\end{definition}

\begin{algorithm}[t]
\footnotesize
\caption{DP control with Restricted and \SDRAshort}
{\bf Input:} $N$: population size; $b$: budget of resources; $\Xbold{0}$: initial infection state; $\mathbf{p}\timeIndex({x})$: transition probability from state $x$ to every other state; $f$: function that gives the number of accessible nodes; $\Lambda\timeIndex$: \pdf of the sample; $\policybold\timeIndex$: \RDRAshort strategy; \emph{isSequential}: specifies if the strategy is \RDRA (false) or \SDRA (true). 
{\bf Output:} $\Xbold{}$: final network state, $\Rbold{}$: final allocation of the resources
\begin{algorithmic}[1]
\State{$\Xbold{} \leftarrow \Xbold{0}$} \algComment{initialize infection state}
\State{$\Rbold{}(\text{randp}(b,N)) \leftarrow 1$} \algComment{initialize resource allocation}
\While{$\ninf{} \neq 0$} 
\State {$n \leftarrow f(\Xbold{})$ } \algComment{compute the number of accessible nodes}
\State {$\Cobold{} \leftarrow \text{find}(\Rbold{} = 1)$} \algComment{currently treated nodes} 
\State {$\Cbold{} \sim \Lambda(n,\Xbold{})$ } \algComment{generate a node sample} 
\If {\text{\emph{isSequential}} == true}
\For {$j =1\mydots n$}\algComment{loop of a selection round}
\State{$\Rbold{} \leftarrow \policybold(\Cobold{}, C_{j})$} \algComment{update resource allocation sequentially}
\EndFor
\Else 
\State{$\Rbold{} \leftarrow \policybold(\Cobold{},\Cbold{})$} \algComment{update resource allocation altogether}

\EndIf
\State{$\Xbold{} \leftarrow \mathbf{p}(\Xbold{})$} \algComment{update infection state}
\EndWhile
\State {\Return $\Xbold{}$, $\Rbold{}$}
\end{algorithmic}
\label{alg:sdra}
\end{algorithm}

The way the \RDRA and \SDRA models operate is described in \Alg{alg:sdra}, and an deployment example is depicted in \Fig{fig:time_scale}.
%%=====================================================
\section{From \DP control to a \MSSPfull}\label{sec:main}
%=====================================================
%-------------------------------------------------------------------------------------------
\subsection{Link with the \SSPfull }
%-------------------------------------------------------------------------------------------
To the best of our knowledge, this work is the first to cast the dynamic \DP control as a problem where decisions are taken as in a \SSPfull (\SSP).
Our goal is to establish this link in the best and most comprehensible way, hence it is beyond our aims to propose a new \SSP setting. Features to consider before chosing an \SSP settings are to have, for instance, single or multiple resources, finite or infinite horizon, score-based or rank-based objective function, \etc

Most SSPs consider a \emph{cold-starting} selection where the \DM begins with an empty selection set. Contrary, in the \SDRA setting and the epidemic control, the resources are supposed to be constantly in effect in the network and, hence, when a selection round starts, there is a \emph{warm-start} where a set of currently treated nodes are in fact already `selected'.
It turns out that this is the biggest difficulty in finding methods in the existing \SSP literature matching the \SDRA setting.
Nevertheless, a \emph{warm-starting} \SSP variant that fits well to the sequential epidemic control has been presented in \cite{Fekom19}.

\begin{figure*}[t]
\centering
{\includegraphics[width=0.90\linewidth, viewport=10 280 1290 770,clip]{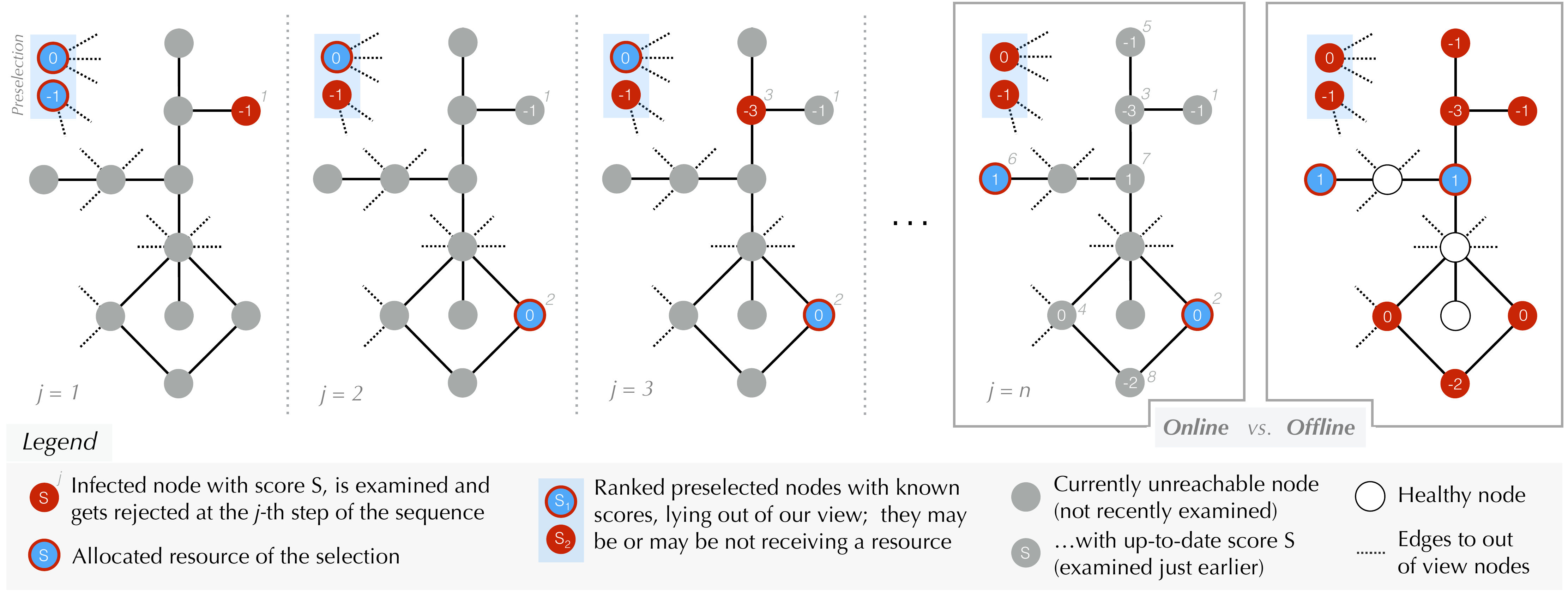}}
\vspace{-0.5em}
\caption{Example of the 3 first steps of an \SDRA round. Candidates, from a sample of $n=8$ infected nodes, appear sequentially \wrt $j$ and possible reallocations are decided immediately. For instance, at step $j=2$, the examined candidate receives a resource that is withdrawn from the second preselected node (lies out of the main graph view). The two rightmost figures compare the result of the online selection to that of the offline selection used by \RDRA.}
\label{fig:online}
\end{figure*}

%---------------------------------------------------------------
\subsection{\WSSPfull}\label{sec:mapping}
%---------------------------------------------------------------
%
We map the problem of \DP control with \SDRA to a succession of separate \emph{\WSSPfull\!\!s} (\WSSP\!\!s) \cite{Fekom19}. 
Specifically, one selection round of the former corresponds to one instance of the latter.
For convenience, in our notations we drop the round subscript $k$  within each \WSSP, \eg $\Cbold{k}$ becomes $\Cbold{}$. 

\begin{definition}{The \emphd{\WSSPfull} (\WSSP)} is an \SSP variant described by elements of different categories: \vspace{1mm}

\noindent1)~\textbf{Background} 
(included in $\mathcal{B}$) 

\noindent a.~Information
\begin{itemize}
\item $b \in \nat$: fixed budget of resources,
\item $s:\nodes \rightarrow \real$: scoring function \st $\Sbold = s(\nodes) \in \real^N$ is the score vector of the entire population,
\item $n\in \nat$: number of candidates to come.

\end{itemize}

\noindent b.~Initialization
\begin{itemize}
\item  $\Cobold{} = (\Co{1}{},\mydots,\Co{b}{}) \subset \nodes$: the subset of the population, called \emph{preselection}, to which resources are initially allocated when a round begins, \ie $\A{\Co{i}{}}{} = 1, \, \forall i \leq b$.
\end{itemize}

\noindent2)~\textbf{Process \& Decisions}
\begin{itemize}
\item $(C_{1},\mydots,C_{n}) \in \Rankset{n}{\nodes \backslash \Cobold{}}$: sequence of randomly incoming candidates for receiving a resource, where $\Rankset{l}{E}$ denotes the set of $l$-combinations of some finite set $E$,
\item $(\A{\cone}{}, \mydots, \A{C_n}{})\in \{0,1\}^n$: sequence of resource allocation decisions taken; giving a resource to a candidate immediately withdraws it from a preselected node (recovered or not), \ie $\A{C_j}{}=1 \Rightarrow \exists\, i\leq b \  \st \ \A{\Co{i}{}}{} = 0$.
\end{itemize}
3)~\textbf{Evaluation}
\begin{itemize} 
\item The cost function is defined as:
\begin{equation} 
\cost_\mathcal{B} = \Sbold \cdot \Abold^{\emph{\text{off}}} - \Sbold \cdot \Abold \, \in \real_+,
\label{eq:cost}
\end{equation}
where $\Abold^\emph{off} = \argmax_{\Abold}\{\Sbold \cdot \Abold \,\,|\,\, R_i = 0, \forall i \notin \access, |\Abold| = b\}$.
\end{itemize}
\label{def:ssp}
\end{definition} 
The first term of the cost function in \Eq{eq:cost} defines the highest achievable score, while the second one gives the score achieved by the taken sequential decisions. As the sequence of incoming candidates is a random variable, $\Exp{\cost_\mathcal{B}/b}$ is the objective function to minimize.

Some observations have to be made concerning our specific \SDRA-to-\WSSP mapping: 1)~It translates the objective of the DP control (\ie to minimize the percentage of infected nodes through time) into an \SSP objective (\ie to minimize the expected cost function of the selected items), hence, $\eta_t = \Exp{\frac{1}{N}\sum_i X_{i,t}}$ is closely related to $\Exp{\cost_\mathcal{B}/b}$. 2)~During each \WSSP instance, the \DM does not need to know anything about the infection state of the network, and merely selects online.
3)~The final selection of a round may not exactly constitute the preselection of the following one, specifically in case of recoveries of nodes that will not be any more part of the preselection. Their resources are returned and become available for the candidates of the round. 4)~If the \DM has still unassigned resources while reaching the end of the sequence of a round (for the triggering function we considered, this can be at most one -- see \Remark{rmk:round-triggering} and below it), then they are by default given to the very last candidates to appear. 
%-----------------------------------------------------------
\subsection{\SSP algorithms for \DP control}
%-----------------------------------------------------------
In \Sec{sec:mapping} we argued about plugging online algorithms from the \SSP literature into the \SDRA model to sequentially control \DPs. Here we focus on the algorithmic part, in particular on two classes of online strategies: 
\begin{itemize}
\item \emph{Cutoff-based}: such a strategy takes as input a \emph{cutoff value} $c\in \natnostar$; it first rejects by default the first $c$ incoming candidates, in a \emph{learning phase}, and then selects a candidate according to information collected during the first phase.
\item \emph{Threshold-based}: a particular case of \emph{cutoff-based} strategies with $c=0$. A candidate is accepted if his score beats a specified \emph{acceptance threshold}.
\end{itemize}
We chose an indicative algorithm from each class: 
the \emph{Cutoff-based Cost Minimization} (\CCM) \cite{Fekom19} and the \emph{Hiring-Above-the-Mean} (\MEAN) \cite{Broder09}, whose objectives are to minimize the expected sum of the ranks (or respectively, the expected sum of scores) of the selected nodes at the end of a round.

\inlinetitle{Cutoff-based Cost Minimization}{.}
\CCM takes also as input a measure of \emph{quality} $q \in ]0,1[$ of the preselected nodes \wrt the sample. This measure indicates how worthy the current resource allocation is. A high quality suggests that the currently treated nodes are among the network's most critical nodes for the epidemic spreading.
For this strategy, the \DM needs to value the selection in terms of ranks instead of scores.

\begin{definition}
The ranking function $\sigma_{N} : \real \times \real^N \rightarrow \{1,...,N\}$ gives to each element of a collection of $N$ values its rank when compared to the other values. 
Let $\mathbf{\Sigma}$ a finite number set, then $\forall s \in \mathbf{\Sigma}:~\!\sigma_N(s, \mathbf{\Sigma}) = \sum_{i=1}^N \Ind{\Sigma_i \le s}$. 
\label{def:ranks}
\end{definition}
Let $\Sobold=(s(\Co{1}{}),...,s(\Co{b})) \subset \Sbold$  be the scores of the candidates and $\mathbf{S^C}= (s(\cone),...,s(C_n)) \subset \Sbold$ the scores of the preselected nodes.
Using \Definition{def:ranks}, the rank-based cost function is defined as $\phi_{\mathcal{B}, k}^{\sigma}=\sum_{S\in \mathcal{S}}  \sigma_{n+b} (S,(\Sobold,\mathbf{S^C}))$, where $\mathcal{S} = \{S_j \in \Sbold | R_j = 1\}_{1\le j\le n}$.
In this paper, we set $q_{k}=\phi_{\mathcal{B}, k-1}^\sigma/b, \, \forall k\leq K$, where $\phi_{\mathcal{B},0}=0.5$. This way, the quality of the preselected nodes in the $k$-th round is simply the sum of their ranks \wrt to the sample of the previous round, when they were selected. This implies our assumption that the item ranks are rather constant between two subsequent rounds. 
Then, the table $c^*(b,n,q) \in \natnostar^{b\times n}$ is computed by tracking the lowest point of the expected rank-based cost provided in \cite{Fekom19}.

The algorithm proceeds as follows. The $b$-best scores recorded during the learning phase of size $c^*$ are stored ordered in a \emph{reference set}. Then, the acceptance threshold starts at the worse score of the set and moves up each time a candidate is accepted, pointing at the next higher score of a non-resigned employee. The process terminates when the end of the sequence is reached. 
For simplicity, we refer to the \CCM strategy with $c=c^*(b,n,q)$ as \CCMstar. 
Note that the rank-based evaluation is particularly suited for the \DP control, as nodes' criticality scores are most likely changing through time.

\inlinetitle{Hiring-above-the-mean}{.}
\MEAN strategy \cite{Broder09} considers the average score of the preselection as acceptance threshold, which is updated after each new selection. 
In the original setting, each incoming candidate $j \le n$ has a \emph{quality score} $Q_j \sim \mathcal{U}(0,1)$; and the goal is to keep a good trade-off between the quality of the selection and speed of hires.
Let the average quality of $b+1$ employees be denoted as $A_b$ (it starts with $A_0 \in [0,1]$). To quantify the rate of convergence of the latter and the rate at which candidates are hired, they define a \emph{gap} $G_b = 1-A_b$, which converges to 0 almost surely when $b$ goes to infinity. Among other results, 
they proved that after $b^{3/2}$ candidate interviews, the expected value for the mean gap for the best $b$ candidates is $O(1/\sqrt{b})$, which makes the strategy close to optimal given this particular evaluation criterion.
Despite being more intuitive and easier to implement than \CCM, this strategy reaches its limit when the preselection is of poor quality with respect to the sample (and probably also with all the population of care-seekers). We can also consider the strategy \cite{Broder09}, where the acceptance threshold used is their median score, \MEDIAN.
%-----------------------------------------------------------
\section{{Offline \vs Online}}\label{sec:off}
%-----------------------------------------------------------

In our \DP context, a strategy is called \emph{offline} when it deterministically selects the $b$-best reachable nodes and immediately assigns resources to them. As explained earlier, the notion of `best' is given by the `black box' scoring function $s:\nodes\rightarrow \real$, which prioritizes nodes independently based on their criticality for the spread. On the other side, an \emph{online} strategy can only examine the candidate nodes one-by-one (see \Definition{def:sdra}). 
Before going further, let us clarify that for simplicity we add the superscript `off' to refer to the offline strategy associated to the online strategy $\mathbf{\Pi}_k(\mathcal{C})$ that is implicitly used. Also, the offline strategy defined in this way is only an indicator of which nodes would have been selected by the oracle.

The main issue that arises from the link with \SSP concerns the relationship between the selection performance of an online strategy and the expected number of infected nodes. This can be rather measured as a difference in the effect at the epidemic compared to the corresponding offline strategy, namely $\diff {\numinfected_t} = \Exp{\numinfected_t - \numinfectedoff_t}$. 
The performance of an online selection strategy is quantified by its expected sum of false negatives (FN) and false positives (FP) in the sequence, hence we define the \emph{online selection error} (or just \emph{error}): $\error_k = \frac{1}{2}\Exp{| \Rbold{k} - {\Rbold{k}}^{\text{off}} |}$. $\error_k$ is evaluated among the subsequent rounds, right after a round's selection is finalized.
An \emph{online strategy with guarantee} can therefore be defined by providing a bound on either:

\noindent i) the expected number of errors over all rounds, \st:
\begin{equation}
{\textstyle \frac{1}{b}\sum_{k=1}^K} \error_k \le M_K,\ \text{with } M_K \in \real_+,
\label{eq:bound1}
\end{equation}
\noindent ii) or the expected cost at any $k$-th round, $\cost_{\mathcal{B},k}$ (see \Eq{eq:cost}), written as $\cost_k = {\Sbold}_k \cdot ({\Rbold{k}}^{\text{off}}-\Rbold{k})$, \st:
\begin{equation}
 \Exp{\cost_k} \le L_k, \,\, \forall k \le K,\ \text{with } L_k \in \real_+.
 \label{eq:bound2}
 \end{equation}
This second bound is stronger as it is given round-wise. However, it requires a certain knowledge of the score distribution.

Going back to the \DP, a thorough numerical analysis lead us to the formulation of the following linear regression assumption.

\begin{assumption} The expected AUC of the difference in the percentage of infected nodes between an online and an offline selection strategy is an affine function of the AUC of the online selection error, which is formulated as:
\begin{equation}
\int_0^T  \frac{\diff \numinfected_t}{N} dt = \cone \left(\sum_{k=1}^K \frac{\error_k}{b}\right) + \ctwo, 
\label{eq:linear_reg}
\end{equation}
where $\cone \geq 0 \geq \ctwo$ are constants that depend on the sampling size $\alpha$ (see \Sec{sec:RDRA}), and the epidemic parameters of the problem.
\label{assum:linear_reg}
\end{assumption}

Suppose that an online strategy provides a bound on its expected number of errors over the rounds and validates \Eq{eq:bound1}. Under the \Assumption{assum:linear_reg}, one can deduce the following bound on the AUC of the percentage of infected nodes:
\begin{equation}
\int_0^T  \frac{\diff \numinfected_t}{N} dt \, \le \,{\cone}   M_K + \ctwo.
\end{equation}
Suppose now that an online strategy provides a bound on the cost at round $k$ and validates \Eq{eq:bound2}. Observe that $\Exp{\cost_k} = \Exp{ {\Sbold}_k \cdot ({\Rbold{k}}^{\text{off}}-\Rbold{k})}\ge \Exp{S_{\min,k}\mathbf{1}_k \cdot ({\Rbold{k}}^{\text{off}}-\Rbold{k}) } = 2\Exp{S_{\min,k}}\Exp{\error_k}$, where $S_{\min,k}$ is the minimum score over all candidates of round $k$.
Thus,  \Eq{eq:linear_reg} in \Assumption{assum:linear_reg} becomes:
\begin{equation}
\int_0^T  \frac{\diff \numinfected_t}{N} dt \, \le \,{\cone}\left( \sum_{k=1}^K \frac{ L_k}{2b\,\Exp{S_{\min,k}}} \right)+ \ctwo.
\end{equation}
A short investigation of the role of the constants $\cone$ and $\ctwo$ can be found in the simulations.

\inlinetitleemph{Example: online \vs offline selection}{ -- }\Fig{fig:online} displays an example of an online selection round where two resources are initially assigned to the nodes of the preselection (top-left in each subfigure). Suppose that the online strategy gives a resource to incoming nodes with score higher than the average score of the preselection, here with scores $\{0,-1\}$ (the higher, the more critical they are). Scores of other infected nodes appear when their turn in the sequence arrives and each of them gets accessed. 
The first candidate ($j=1$) is not selected since his score of $-1$ does not beat the threshold of $\frac{1}{2}(-1+0)=-0.5$. The second candidate ($j=2$), though, gets the resource unit from the worse preselected node. The new score threshold to beat is set to $0$. The process continues, up to the last candidate ($j=n$). The final resource allocation of an offline selection strategy is also shown (rightmost subfigure). Here, the cost function for the online case is $\frac{1}{b}\cost_\mathcal{B} = (1+1)-(1+0)=1$, where the first term is the highest achievable average score of the selection (\ie the offline score), and the second term is what the online strategy achieved. 
Regardless the scoring function, an efficient \SDRA strategy (online) should be as close as possible to the associated \RDRA strategy (offline) in terms of $A_N(T)$ (see \Eq{eq:auc}).
%==========================================================
\section{Simulations}
%==========================================================
\subsection{Experimental setup}
\begin{figure*}[t]
\centering
\includegraphics[width=1\linewidth, viewport=21 420 1690 745, clip]{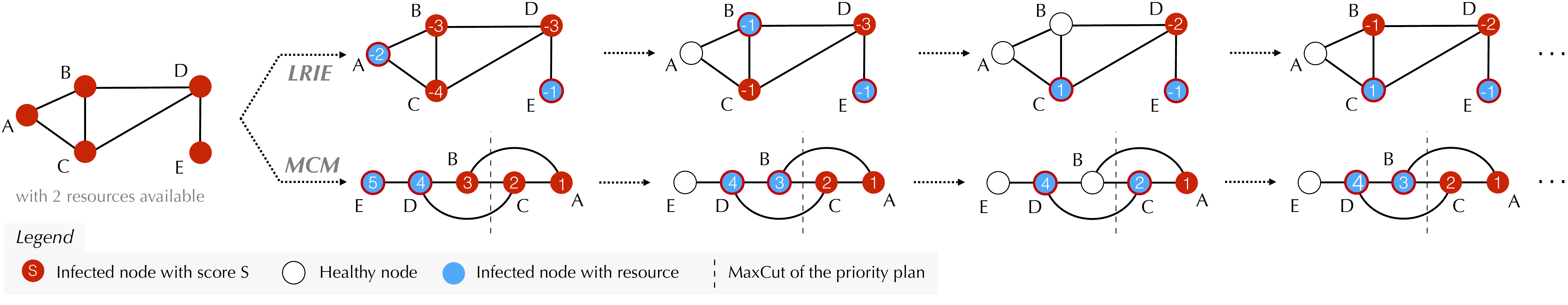}
\vspace{-1.5em}
\caption{Example of the evolution of the network infection state when considering the dynamic scoring function LRIE and the static priority plan of MCM.}
\label{fig:MCMvsLRIE}
\vspace{-1mm}
\end{figure*}
%---------------------------------------------------------
\inlinetitle{Network}{.}
%---------------------------------------------------------
The interactions among a population of $N$ individuals are modeled by a fixed, symmetric (undirected), and unweighted network with adjacency matrix $\adjbold \in \{0,1\}^{N\times N}$ where $\adj_{ij}=1$ only if nodes $i$ and $j$ are linked with an edge.
The connectivity structure is generated according to a \emph{scale-free} (SF), a \emph{small-world} (SW), a \emph{community-structured} (CS) network model, or a real network of Facebook user-user friendships.
\begin{itemize}
\item In the SF type, the node degree distribution follows a power law, hence few nodes are \emph{hubs} and have much more edges than the rest. We use the Barab\'asi-Albert \emph{preferential attachment} model \cite{Barabasi99} that starts with two connected nodes and, thereafter, connects each new node to $m\in \nat$ existing nodes, randomly chosen with probability equal to their normalized degree at that moment.
\item In the SW type, nodes are reachable to each other through short paths. We use the Watts-Strogatz model \cite{Watts98}: it starts by arranging the $N$ nodes on a ring lattice, each connected to $m\in \nat$ neighbors, $m/2$ on each side. Then, with a fixed probability $p\in [0,1]$ for each edge, it decides to rewire it to a uniformly chosen node of the network. 

\item In the CS type, nodes are grouped into sets that are densely connected internally and sparsely between groups. We use an hierarchical \Erdos model, where the probability $p_l\in [0,1]$ for creating each edge reduces at each level $l$ as we move up in the hierarchy. At the lowest level there are $12$ groups of $100$ nodes, $N=1200$ nodes overall (see \Fig{fig:community}).
\end{itemize}

\smallskip
For the two first types of generated networks, we use a small population size of $N=100$ individuals, which however is sufficient for our demonstration. Furthermore, by rescaling the epidemic parameters, the same phenomena can be reproduced in larger networks.
The model parameters to generate each network, are mentioned explicitly in the associated figures.
%---------------------------------------------------------
\inlinetitle{Scoring function}{.}
%---------------------------------------------------------
In this section, we briefly expose a series of known scoring functions that we used in our experiments:
\begin{itemize}
\item \emph{Random} (RAND): selects nodes uniformly at random, among the infected nodes. 
\item \emph{Largest Reduction in Spectral Radius (LRSR)}: selects each time the infected nodes that lead to the largest drop in the first eigenvalue of the network's adjacency matrix \cite{Tong12}.
\item \emph{Largest Reduction in Infectious Edges} (LRIE) \cite{Scaman15}: selects the infected nodes that minimize the number of infectious edges that connect an infected and a healthy node. The associated score is the difference between the number of healthy and infected neighbors of each node:
\begin{equation}
S_{i,t} = {\textstyle\sum_j} \left(\adj_{ij} \bar{X}_{j,t} - M_{ji} X_{j,t} \right),
\label{eq:lrie}
\end{equation}
recall that $\bar{X}_{i,t} = 1 - X_{i,t}, \, \, \forall j, t$.
This scoring function is derived from the minimization of the second order derivative of the expected number of infected nodes \wrt time: $\frac{d^2 \Exp{\numinfected_t}}{dt^2}$.
LRIE is greedy and dynamic, since node scores change each time the infection state and/or the network structure changes.
\item \emph{Max-Cut Minimization} (MCM) \cite{Scaman16}: A \emph{priority planning} strategy computes a \emph{healing plan} which is an ordering of the nodes that accounts for their criticality \wrt the epidemic spread, and that is given prior to the beginning of the diffusion. It is thereby a static scoring function. The resources are given to the first $b$ nodes of the plan, and once those nodes get cured, the resources are reallocated to the next nodes of the plan. 
The MCM finds a node priority-order with as low as possible max-cut (see \Definition{def:maxcut}).
For a network with adjacency matrix $\adjbold = (\adj_{ij}) \in \{0,1\}^{N\times N}$\!, the max-cut of a given priority-order $\ell$ is defined as:  
\begin{equation}
CUT(\ell) = \underset{c=1,...,N}{\max} {\textstyle\sum_{i,j}} \adj_{ij} \mathbbm{1}_{\ell(i) < c < \ell(j)},
\label{def:maxcut}
\end{equation}
where $\ell(i)$ is the order of node $i$ in the plan. Finally, the scoring functions of the infected nodes are given by:
\begin{equation}
 S_{i} = N +1 - \ell(i), 
\label{eq:mcm}
\end{equation}
\ie priority is given to nodes at the beginning of the ordering.
Using such a strategy, and under some specific assumptions, a bound over the expected extinction time can be retrieved, defined as $t_{\text{exct.}}= \min (t, \Xbold{t} = 0)\, \in \real_+$.
\end{itemize}

\inlinetitleemph{Example}{ -- }%
In \Fig{fig:MCMvsLRIE} is displayed the early evolution of the diffusion over the graph $\Graph$ starting with full infection, \ie $\Xbold{0}= (1,...,1)^\top$. The \DM manages dynamically $b=2$ resources (blue nodes).
Following the LRIE strategy, node scores have negative initial values due to the full infection, which however increase as more nodes recover.
We can see that the two highest LRIE scores are spread across the network, as in each score computation only local information is taken into account. 
On the other side, following the MCM strategy implies to first compute the priority-order $\ell=\{E,D,B,C,A\}$ that minimizes the max-cut, which is $CUT(\ell) = 3$ (between nodes $B$ and $C$) in this case, and provides fixed node scores. 

%---------------------------------------------------------
\inlinetitle{Diffusion process and score-based DRA}{.}
%---------------------------------------------------------
The diffusion process we simulate in our experiments is as described in \Sec{sec:SIS-DRA} and with a fixed budget of $b=5$ resources.
In this SIS formulation we have dropped the self-recovery (\ie $\delta = 0$ in \Eq{eq:rates}) in order to emphasize the role of the decisions taken by the compared strategies.
Concerning the scoring function, in the simulations we use both a static and a dynamic scoring function, namely MCM and LRIE.
%---------------------------------------------------------
\subsection{Comparing online strategies}\label{sec:online_comp}
%---------------------------------------------------------
\begin{figure}[t]
\centering
\vspace{-1mm}
\hspace{-3mm}
\subfigure[{\scriptsize Cutoff-based \SDRA on SW}]
{
\clipbox{0pt 0pt 0pt 0pt}
{\includegraphics[width = 0.51\linewidth, viewport=0 130 580 690, clip]{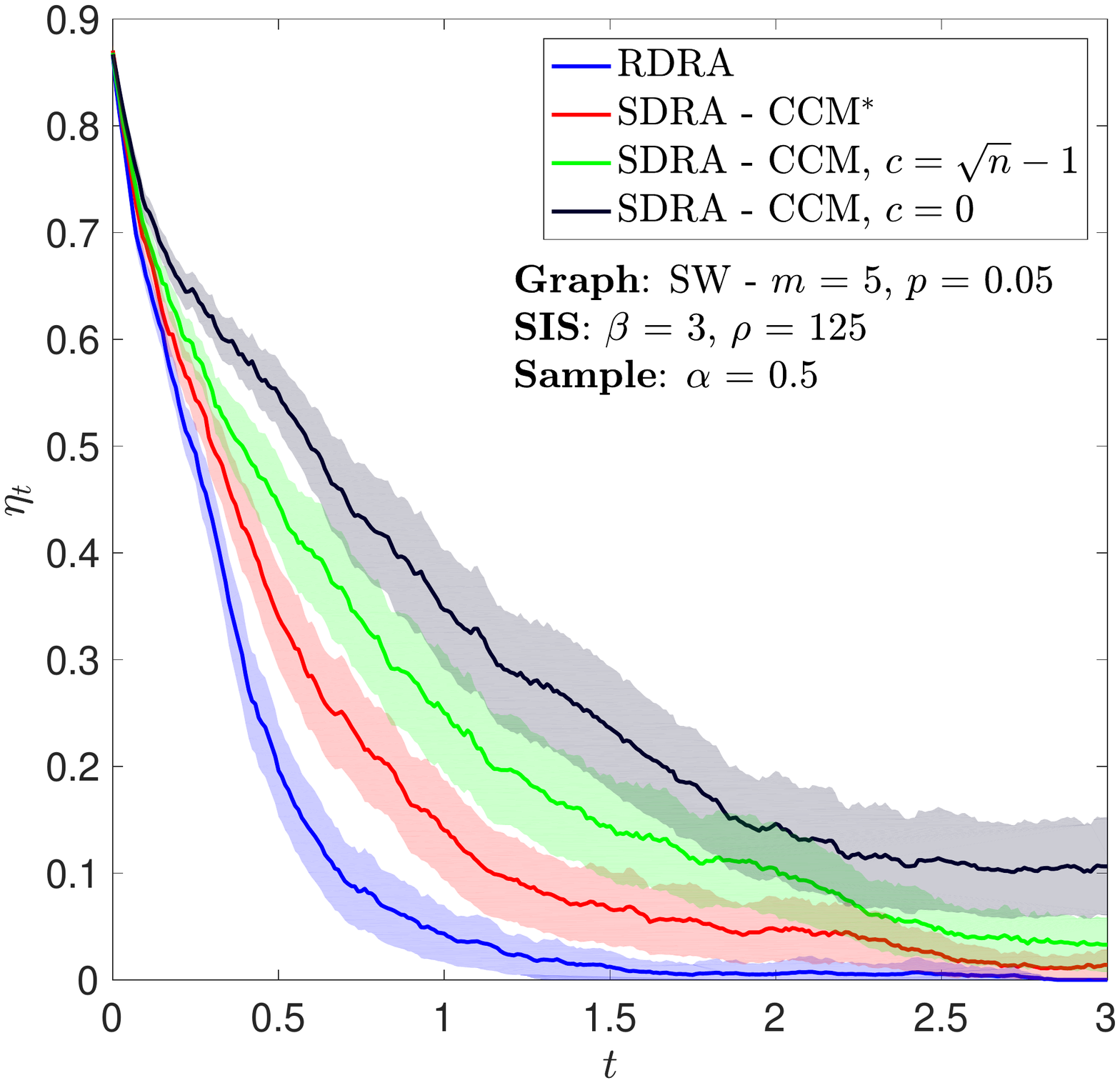}}\label{fig:sw_perc_inf_left}}
\vspace{2mm}
\hspace{-3mm}
\subfigure[{\scriptsize Threshold-based \SDRA on SW}]
{
\clipbox{5pt 0pt 0pt 0pt}
{\includegraphics[width = 0.51\linewidth, viewport=0 130 580 690, clip]{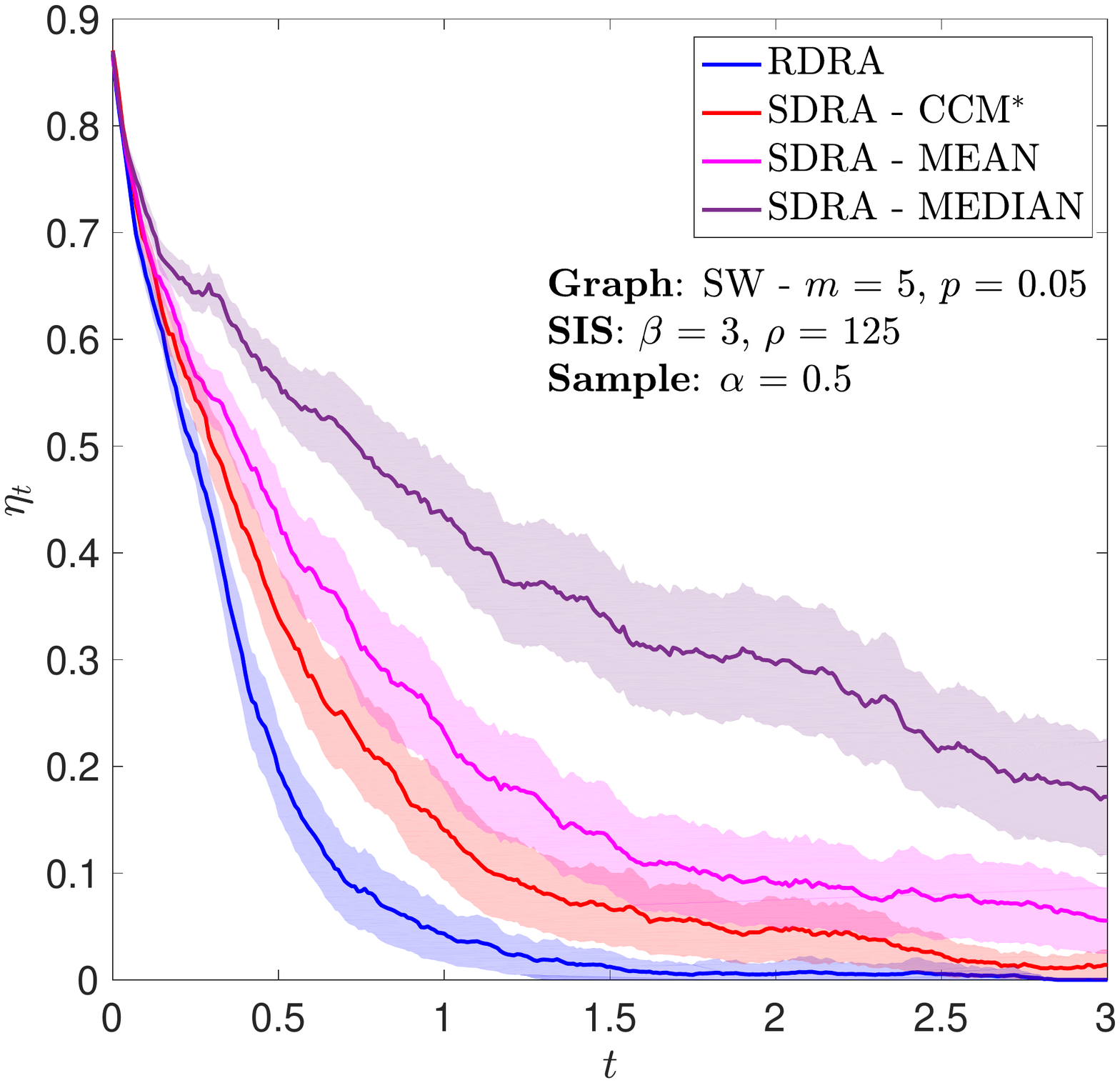}}\label{fig:sw_perc_inf_right}}\\
\vspace{-3mm}
\hspace{-3mm}
\subfigure[{\scriptsize Cutoff-based \SDRA on SF}]{
\clipbox{0pt 0pt 0pt 0pt}
{\includegraphics[width = 0.51\linewidth, viewport=0 130 580 690, clip]{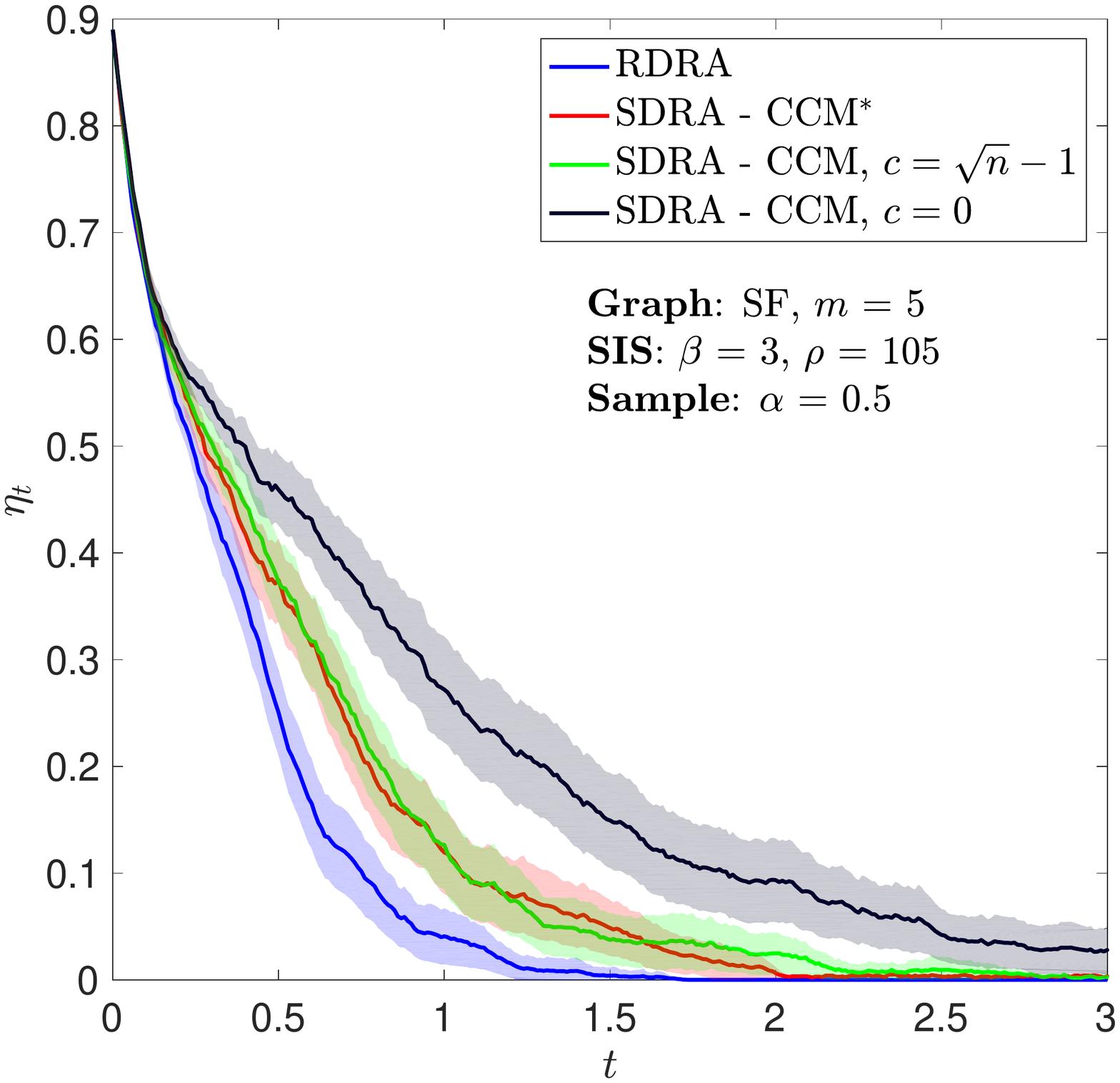}}\label{fig:pa_perc_inf_left}}
\hspace{-3mm}
\subfigure[{\scriptsize Threshold-based \SDRA on SF}]{
\clipbox{5pt 0pt 0pt 0pt}
{\includegraphics[width = 0.51\linewidth, viewport=0 130 580 690, clip]{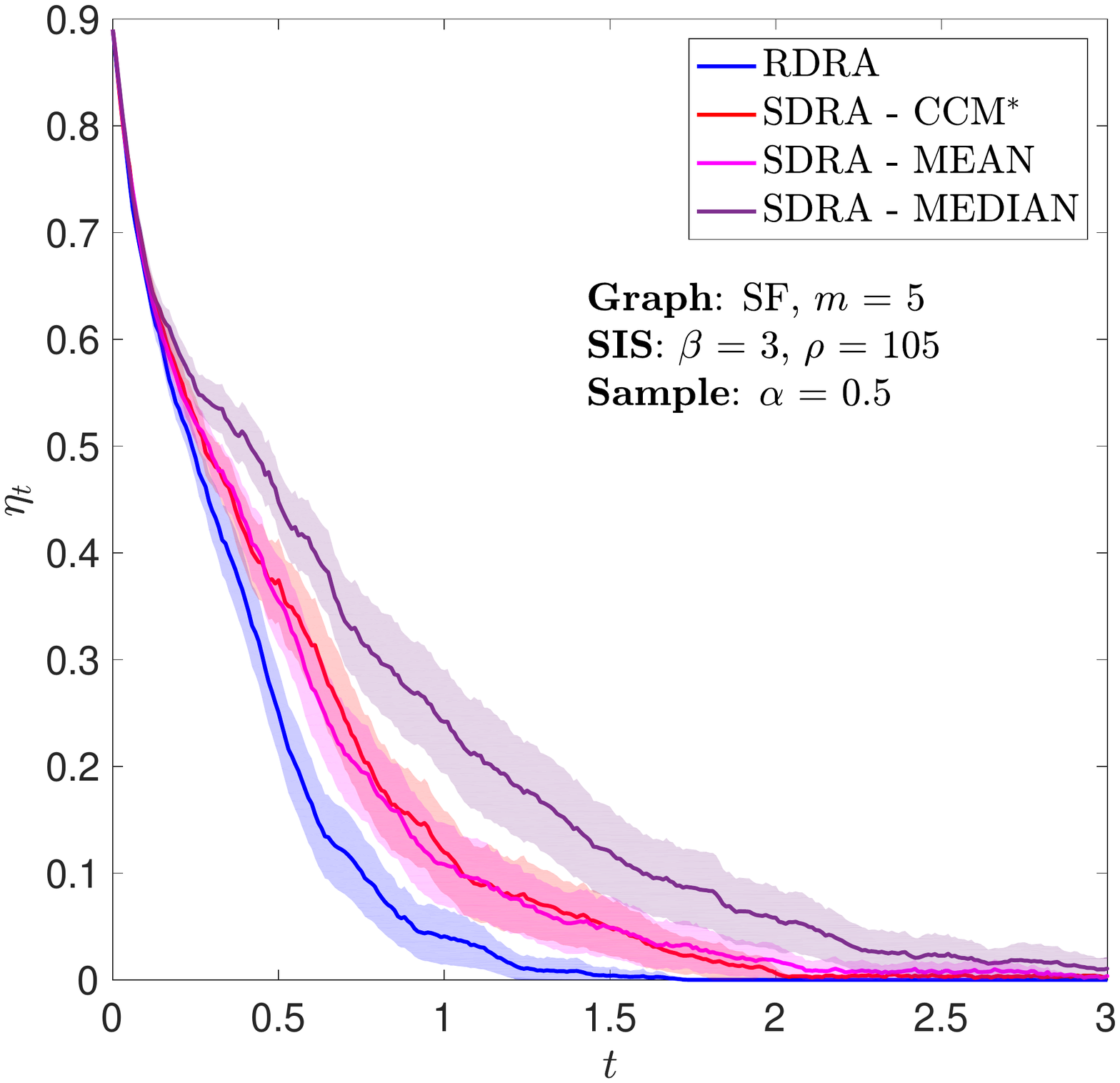}}\label{fig:pa_perc_inf_right}}\\
\caption{Comparison of cutoff-based and threshold-based SDRA strategies, in terms of the average percentage of infected nodes $\eta_t$ through time, for SW (top row) and for SF (bottom row) networks. The RDRA is shown for reference; for the same reason, the proposed SDRA-CCM is also repeated in the right subfigure of each row.}
\label{fig:perc_inf}
\end{figure}
\vspace{-1mm}

In the empirical study, our aim is to compare the performance of several DRA strategies that follow \Alg{alg:sdra}. 
The offline selection strategy, which picks the reachable candidates with the highest LRIE (resp. MCM) scores, is always plotted with a dark blue (resp. light blue) curve as reference (see \Figs{fig:perc_inf},\,\ref{fig:score_functions},\,\ref{fig:perc_inf_mcm}). Other colors imply the sequential allocation of resources. At each round, a fraction $\alpha\in[0,1]$ of the infected nodes $n_t = \lfloor \alpha \sum_iX_{i,t}\rfloor$ is uniformly sampled and become accessible to the \DM.

\Fig{fig:perc_inf} displays the evolution of the average percentage of infected nodes with time, using the compared DRA strategies on two network types. 
The SW type at the top row, where on the left appears the cutoff-based \algo strategy with various cutoffs, and on the right the \MEAN and \MEDIAN variations of the threshold-based strategy. In both subfigures, \CCMstar is clearly the best performing approach. Here, \MEAN is a lot better than \MEDIAN, however, on an SF network (bottom row), the curves appear to be closer together and they do not differ in performance. The \CCM is still better, but \CCMstar shows no improvement over the use of the simpler cutoff $c = \sqrt{n} - 1$.

\begin{figure}[] 
\centering
\vspace{-1mm}
\hspace{-5mm}
\subfigure[{\scriptsize \RDRA on SW}]{
\clipbox{0pt 0pt 0pt 0pt}
{\includegraphics[width = 0.52\linewidth, viewport=12 15 600 570, clip]{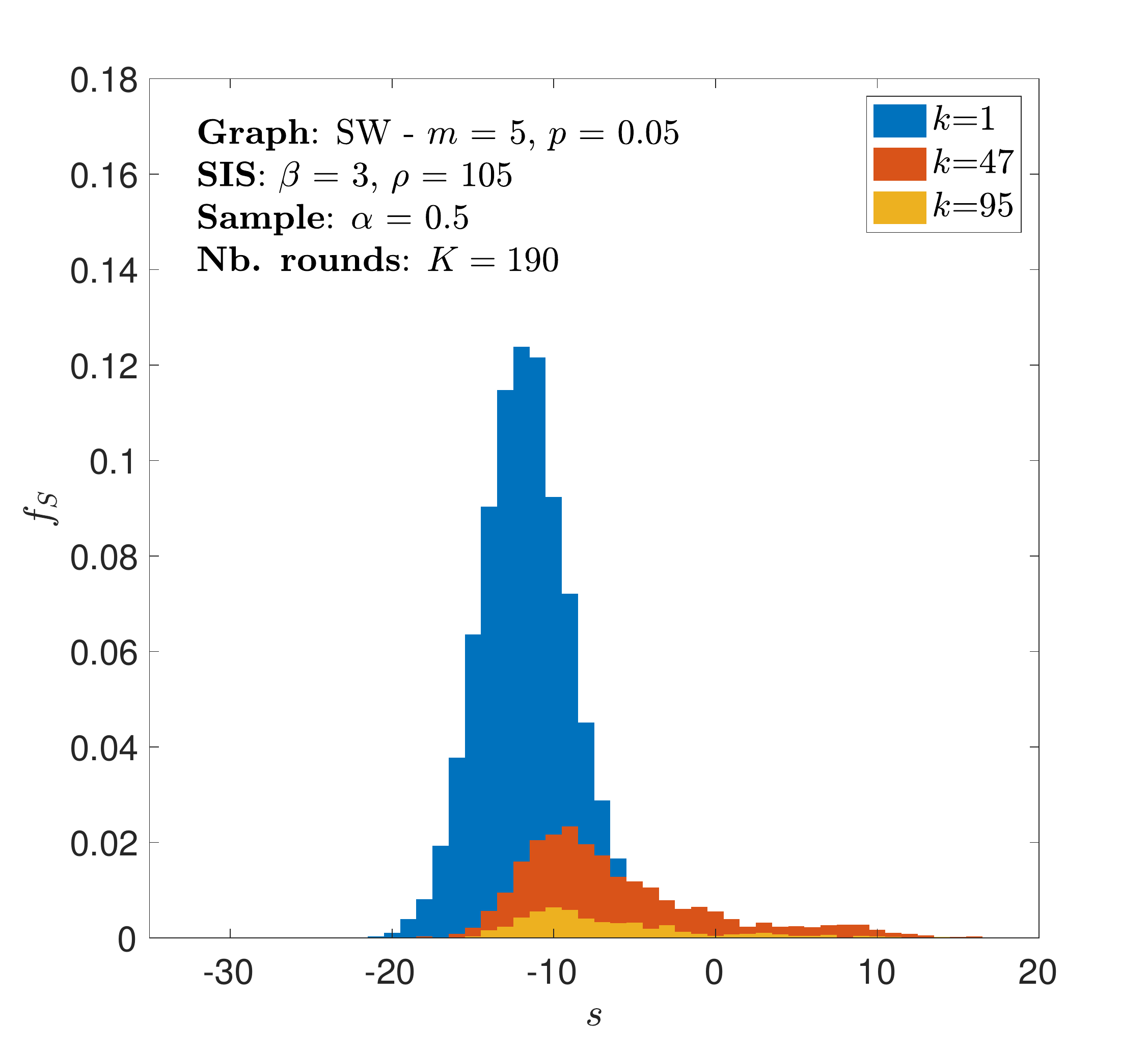}}\label{fig:sw_score_distrib_left}} 
\hspace{-3mm}
\subfigure[{\scriptsize \SDRA, \CCMstar on SW}]{ 
\clipbox{6.4pt 0pt 0pt 0pt}
{\includegraphics[width = 0.52\linewidth, viewport=12 15 600 570, clip]{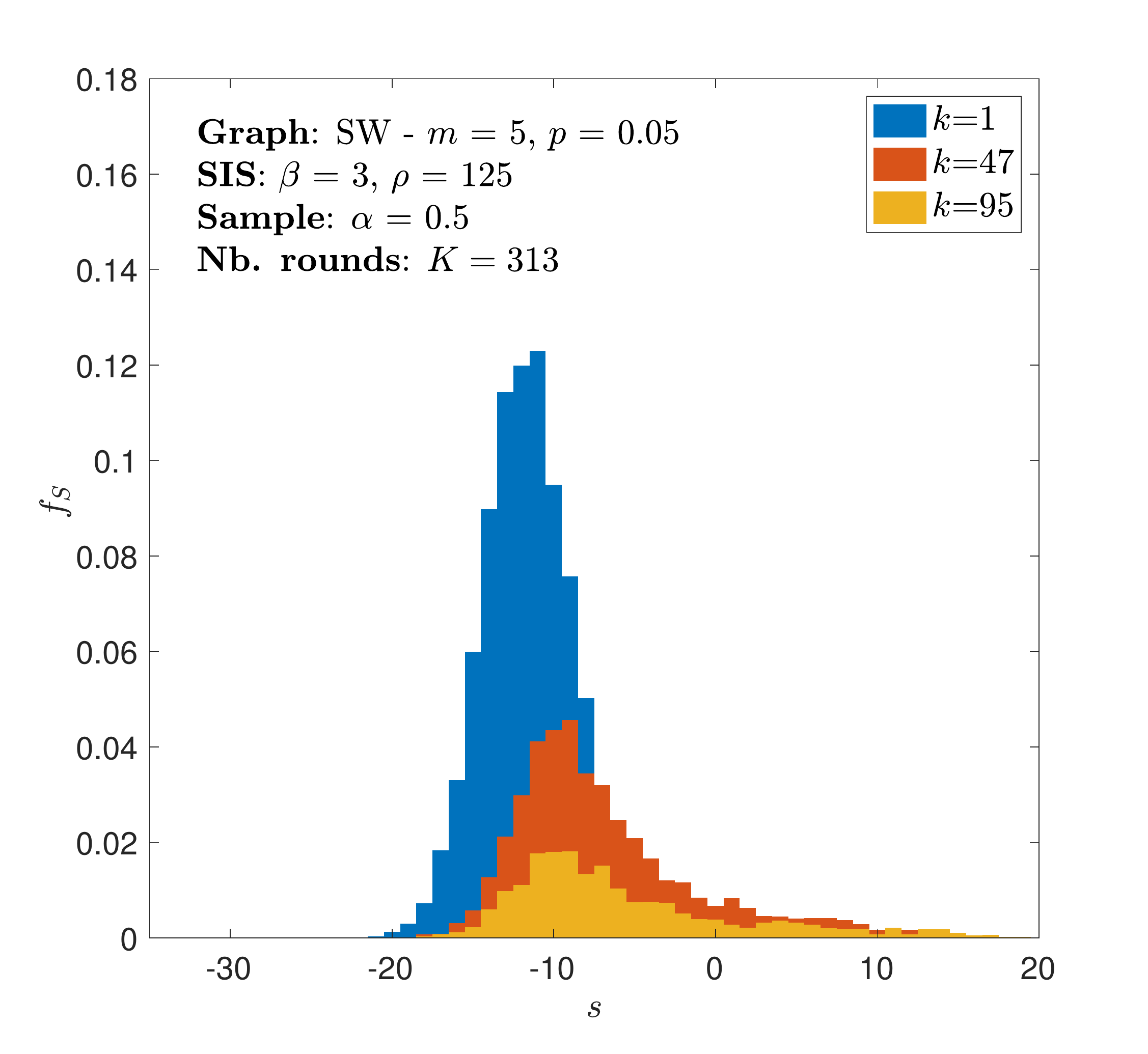}}\label{fig:sw_score_distrib_right}}\\
\vspace{-1mm}
\hspace{-4.2mm}
\subfigure[{\scriptsize \RDRA on SF}]{
\clipbox{0.8pt 0pt 0pt 0pt}
{\includegraphics[width = 0.52\linewidth, viewport=12 15 600 570
, clip]{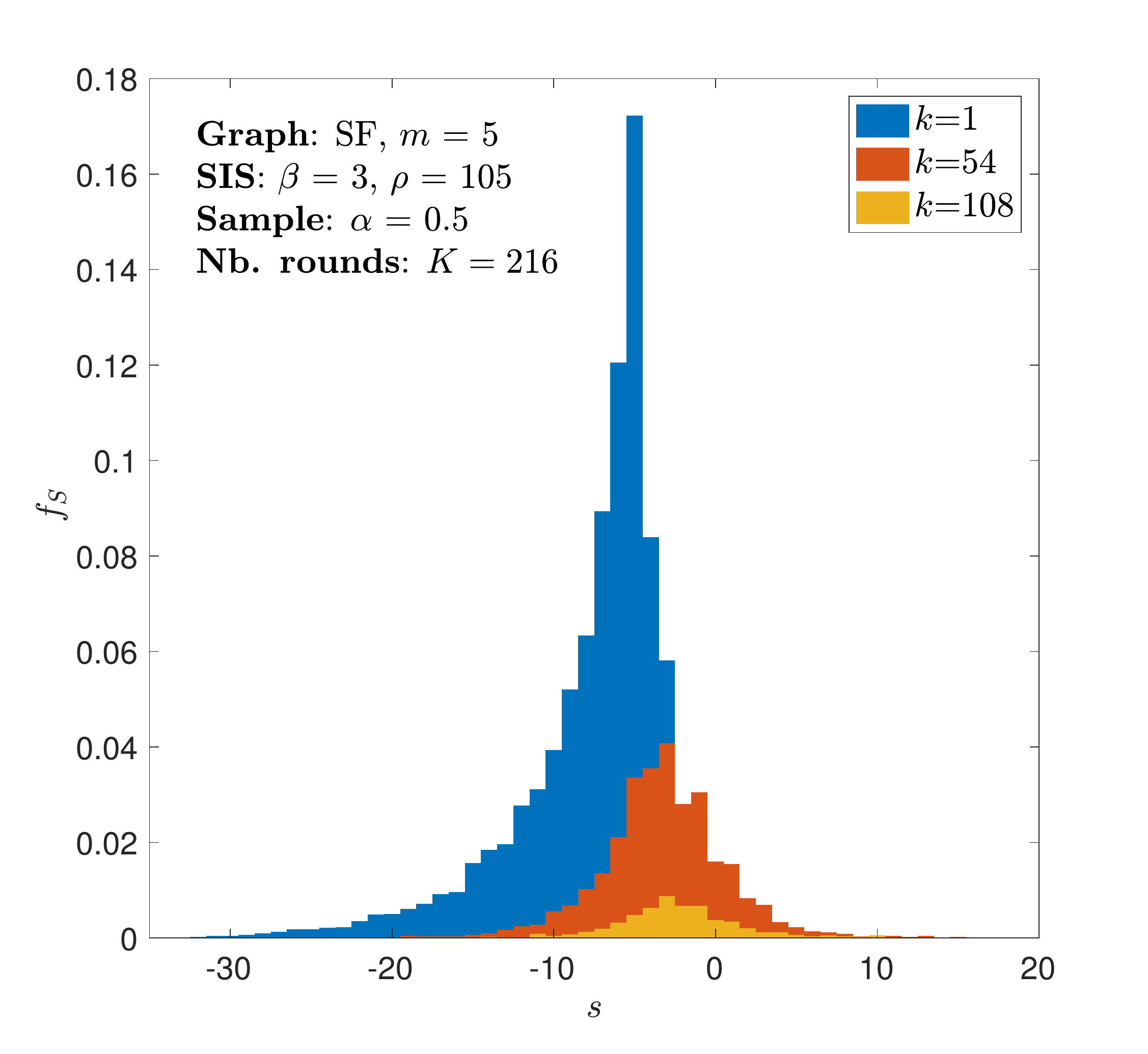}}\label{fig:sf_score_distrib_left}}
\hspace{-3mm}
\subfigure[{\scriptsize \SDRA, \CCMstar on SF}]{ 
\clipbox{6.8pt 0pt 0pt 0pt}
{\includegraphics[width = 0.52\linewidth, viewport=12 15 600 570, clip]{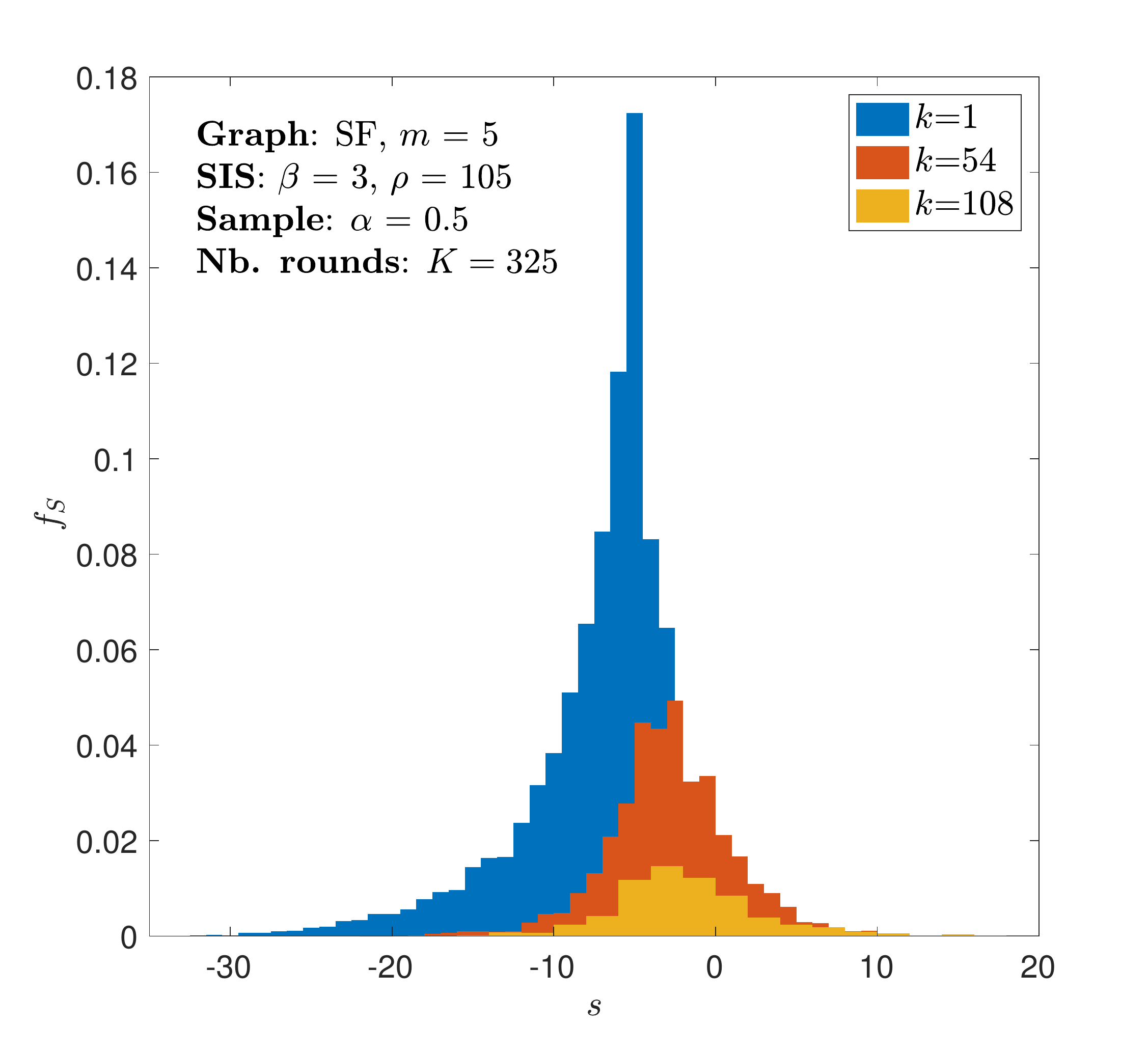}}\label{fig:sf_score_distrib_right}}
\vspace{-2.1mm}
\caption{The empirical \pdf $f_S$ of the node scores, at different different rounds $\t$ through the multi-round process, for SW (top row) and for SF (bottom row) networks. The results refer to the application of \RDRA or \SDRA (the \CCMstar) strategies with the LRIE scoring function (initialization with the same infection level in each row).}
\label{fig:score_distrib}
\vspace{-1.0mm}
\end{figure}

To further investigate the behavior of the strategies, we plot in \Fig{fig:score_distrib} the score distribution $f_S$ (here from LRIE) for the infected nodes of the network at three different rounds in the course of a multi-round process. The top and bottom rows refer respectively to SW and SF networks. \Fig{fig:sw_score_distrib_left} and \Fig{fig:sf_score_distrib_left} show the $f_S$ obtained using a \RDRA strategy (blue curves in \Fig{fig:perc_inf}), while for \Fig{fig:sw_score_distrib_right} and \Fig{fig:sf_score_distrib_right} the sequential strategy used is the \CCMstar (red curve in \Fig{fig:perc_inf}). Starting from nearly identical $f_S$ per row at $k=1$ (same initial infection level), we observe that throughout the rounds the difference between the distributions of the \RDRA and \SDRA strategies is larger for SW networks. This is as expected since in that example the strategies have larger difference in performance. Moreover, for SF networks, the $f_S$ leans towards a Gaussian-like shape, which explains why \MEAN and \MEDIAN behave similarly, contrary to the more skewed shape obtained for a SW network.

It is easy to see in these simulations that the network structure plays a crucial role in the epidemic spread and sets the difficulty level to any strategy that tries to contain it. Also, the highly evolving shape of the score distribution throughout the process (rounds) illustrates the challenges that SDRA strategies need to address in order to be sufficiently effective. 

%--------------------------------------------------------------------------------
\subsection{\RDRA with different scoring functions}
\begin{figure}[t]
\vspace{-1mm}
\centering
\hspace{-3mm}
\subfigure[$\alpha = 1$]{
\clipbox{0pt 0pt 2pt 2pt}
{\includegraphics[width = 0.51\linewidth, viewport=0 130 580 690, clip]{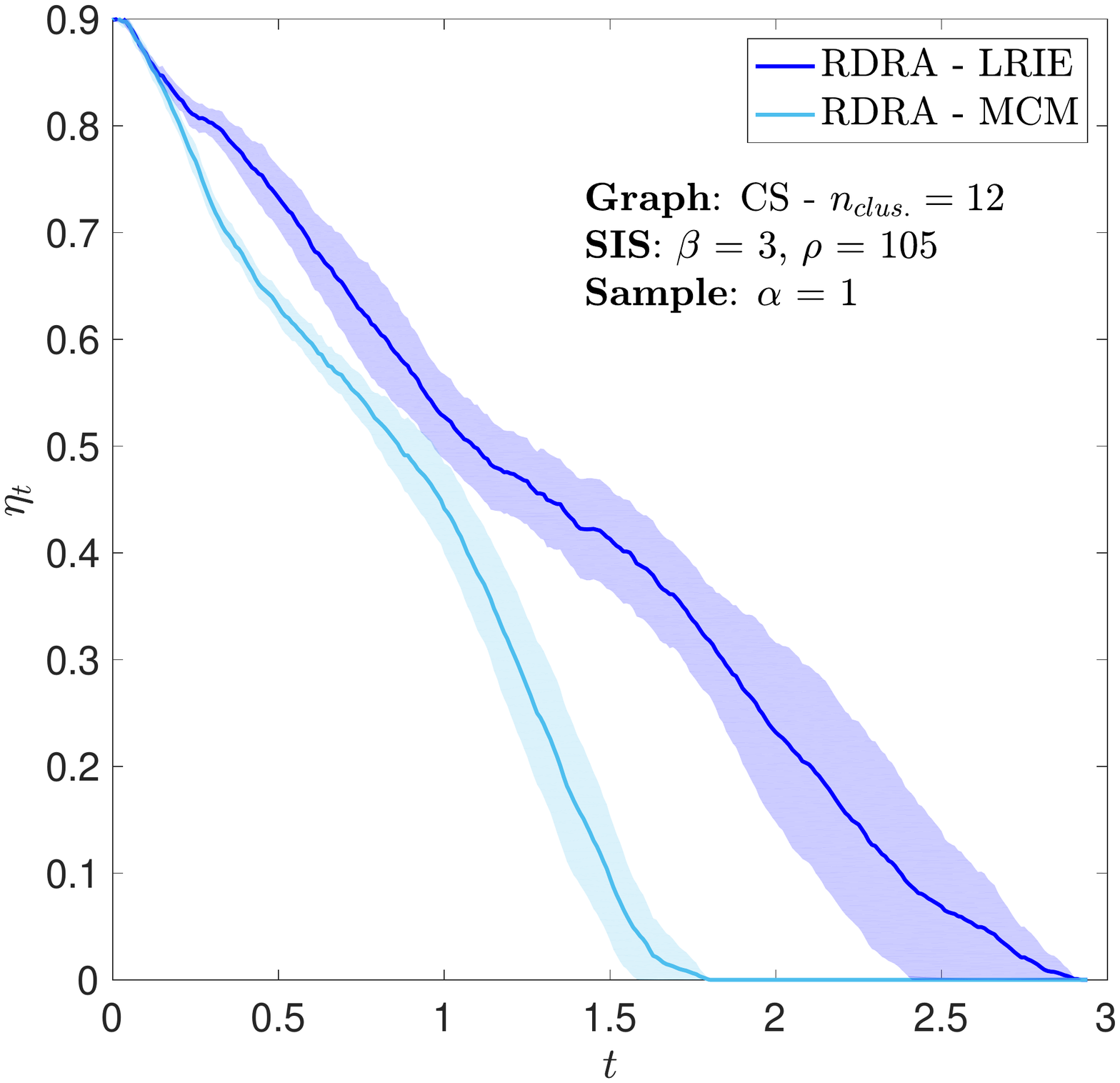}}\label{fig:alpha1}}
\hspace{-2mm}
\subfigure[$\alpha = 0.05$]{ 
\clipbox{4pt 0pt 2pt 2pt}
{\includegraphics[width = 0.51\linewidth, viewport=0 130 580 690, clip]{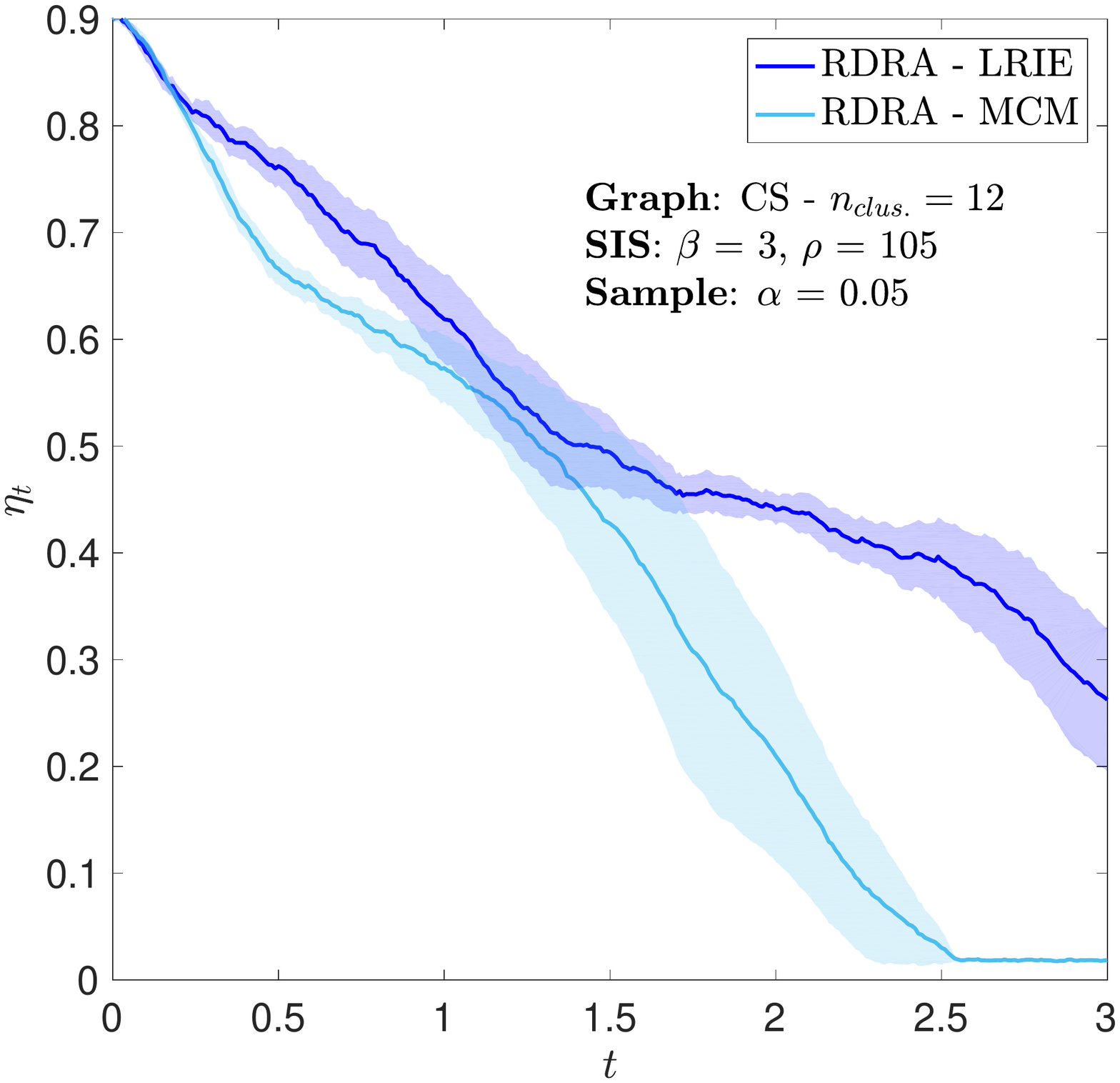}}\label{fig:alpha005}}\\ 
\vspace{-2mm}
\caption{Comparison of the performance of the \RDRA with LRIE (dark blue) and MCM (light blue). Average percentage of infected nodes \wrt time $t$, $\eta_t$ for a CS network of $N=1200$ nodes, 4 high-level groups, and 12 low-level groups, displayed in \Fig{fig:community}. The number of resources is $b=17$.}
\label{fig:score_functions}
\vspace{-1mm}
\end{figure}

%--------------------------------------------------------------------------------
\inlinetitle{CS network}{.}
So far we have used the LRIE scoring function for the prioritization of infected nodes. Here, we compare the behavior of the \RDRA strategy when using either one of the dynamic LRIE or the static MCM functions. 
In \Fig{fig:score_functions}, the percentage of infected nodes \wrt time is displayed for a CS network that exhibits a hierarchical community structure, see \cite{Allen11}. Due to the large variation of edge density that such networks exhibit, they usually require less resources than a graph with the same number of edges but without community structure; in this case $b=17$ resources are enough for $N=1200$ nodes.
We notice that \RDRA performs better with the MCM scoring function than with the LRIE, as it is more efficient at targeting the critical nodes.
\Fig{fig:score_functions} shows the impact of the sampling size (left: $\alpha = 1$; right: $\alpha = 0.05$). Despite the significant difference between the two sampling sizes, the efficiency is only slightly reduced from one to the other. 

Overall, in practice our framework seems to be able to translate the quality of a scoring function to better performance in the constrained RDRA setting, and its applicability remains high even when the sampling size is quite small. 

\begin{figure}[t]
\centering
\vspace{-0.9em}
\hspace{-3mm}
\subfigure[{\scriptsize Cutoff-based \SDRA on SW}]
{
\clipbox{0pt 0pt 3pt 0pt}
{\includegraphics[width = 0.51\linewidth, viewport=0 130 580 690, clip]{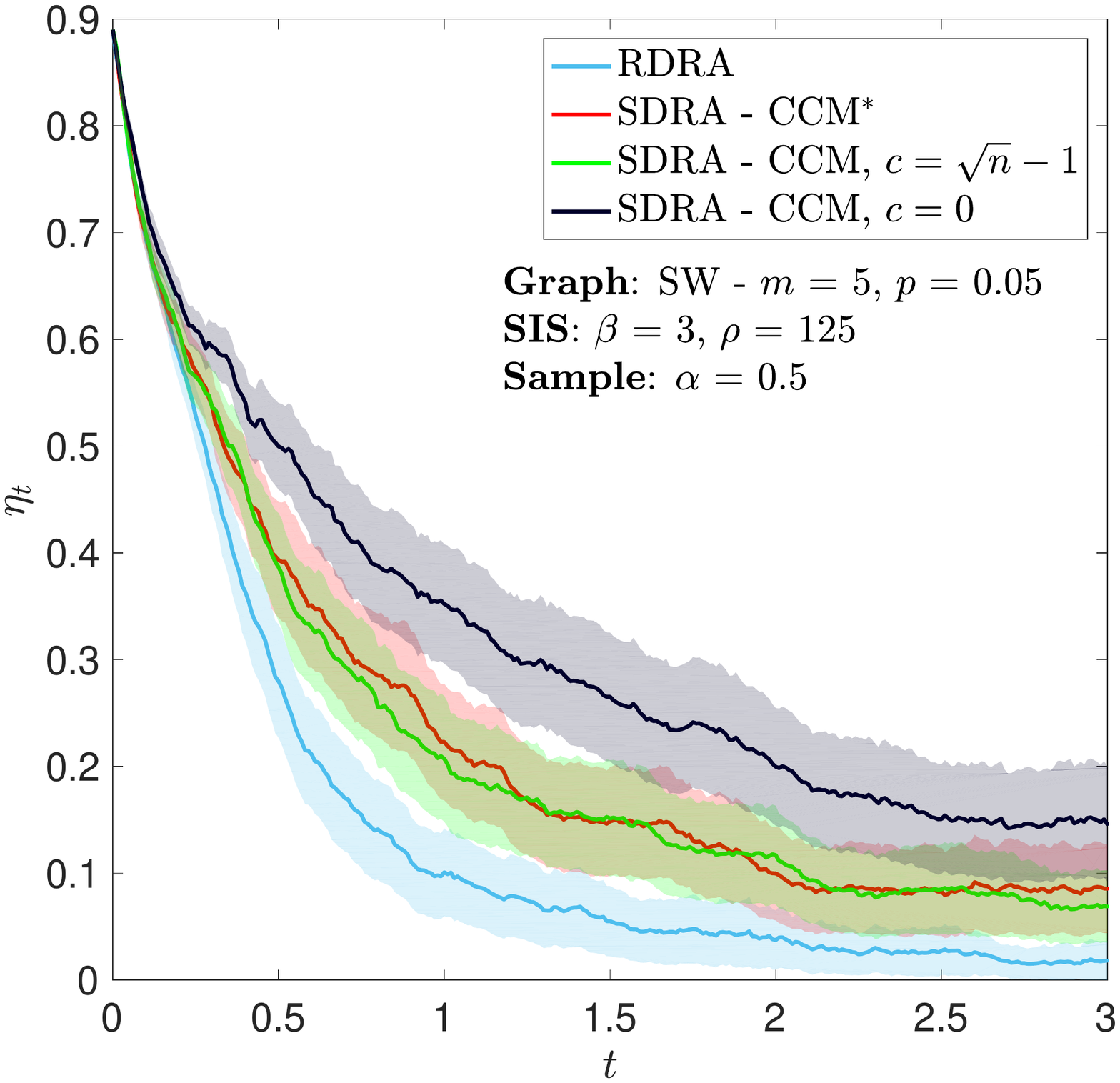}}\label{fig:sw_perc_inf_left_MCM}}
\vspace{1mm}
\hspace{-1mm}
\subfigure[{\scriptsize Threshold-based \SDRA on SW}]
{
\clipbox{5pt 0pt 3pt 0pt}
{\includegraphics[width = 0.51\linewidth, viewport=0 130 580 690, clip]{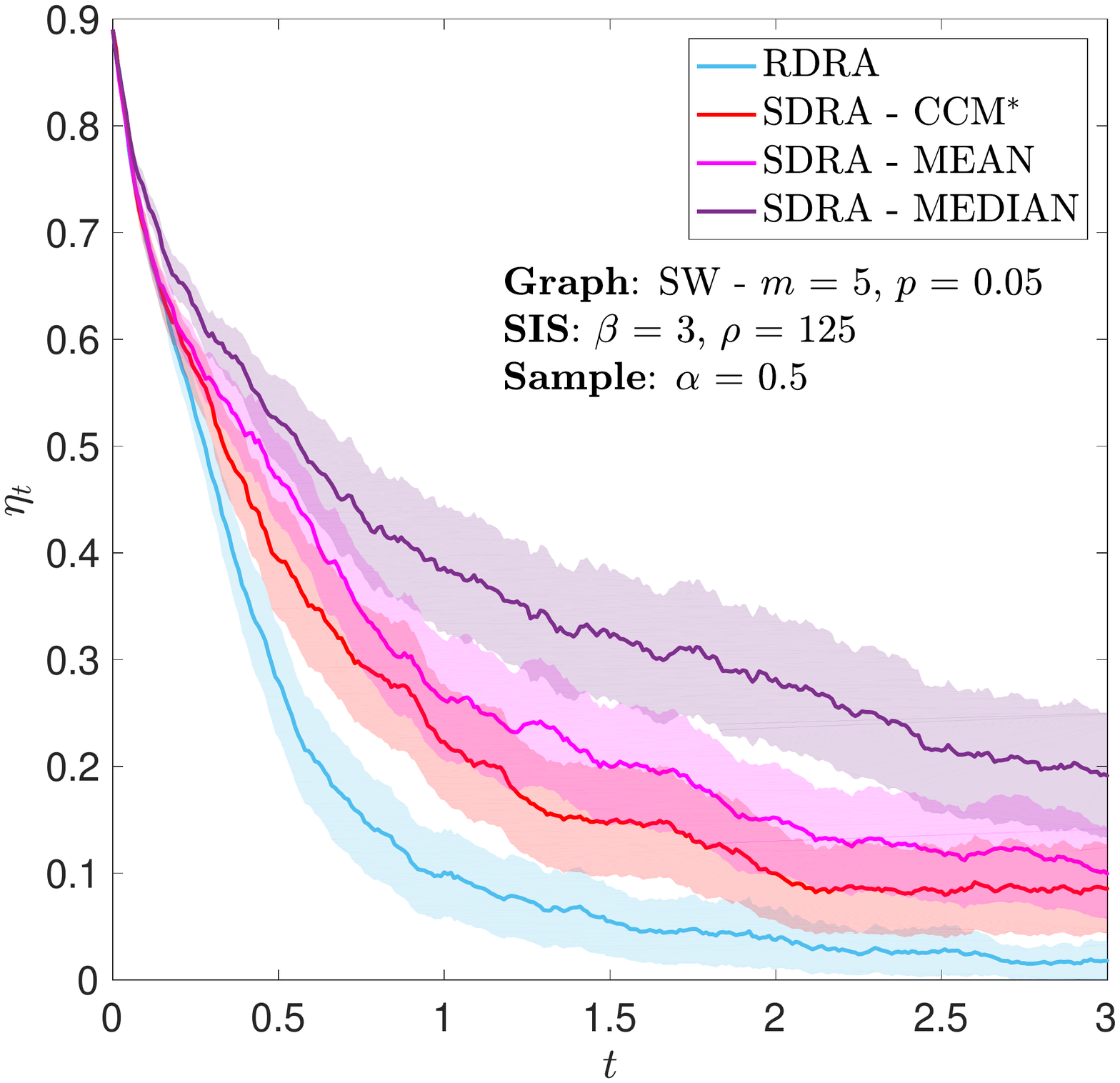}}\label{fig:sw_perc_inf_right}}\\
\vspace{-3mm}
\hspace{-3mm}
\vspace{-2mm}
\caption{Comparison of cutoff-based and threshold-based SDRA strategies. The Restricted DRA is shown in light blue for reference; for the same reason, the proposed SDRA-CCM is also repeated in the right figure. Average percentage of infected nodes $\eta_t$ through time for SW network.}
\label{fig:perc_inf_mcm}
\vspace{-1mm}
\end{figure}
\begin{figure}[t] 
\vspace{-1mm}
\centering
\hspace{-4mm}
\subfigure[{\scriptsize \RDRA on SW}]{
\clipbox{0pt 0pt 2pt 2pt}
{\includegraphics[width = 0.532\linewidth, viewport=5 15 620 580, clip]{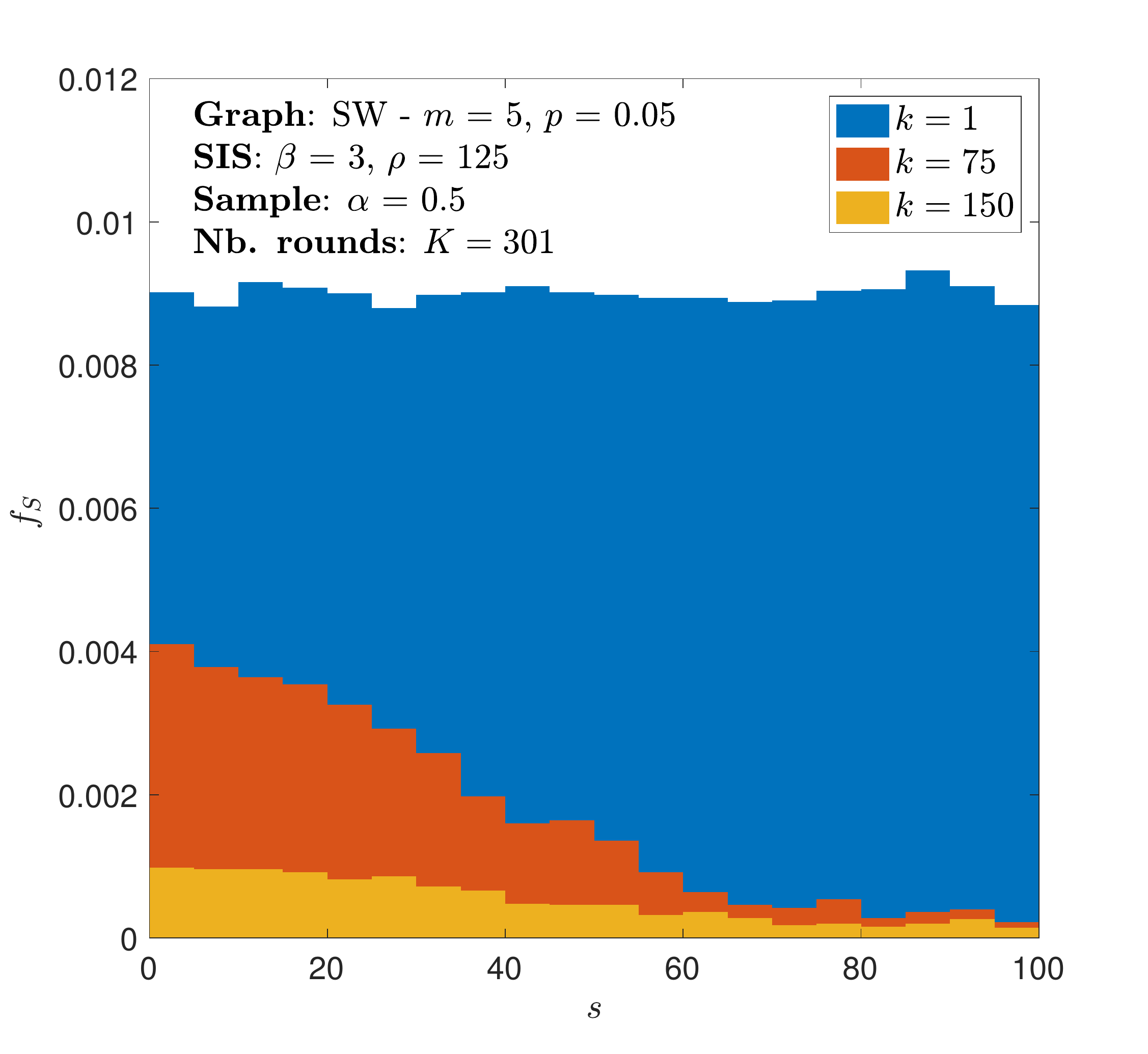}}\label{fig:---}}
\hspace{-4.0mm}
\subfigure[{\scriptsize \SDRA, \CCMstar on SW}]{ 
\clipbox{6pt 0pt 2pt 2pt}
{\includegraphics[width = 0.532\linewidth, viewport=5 15 620 580, clip]{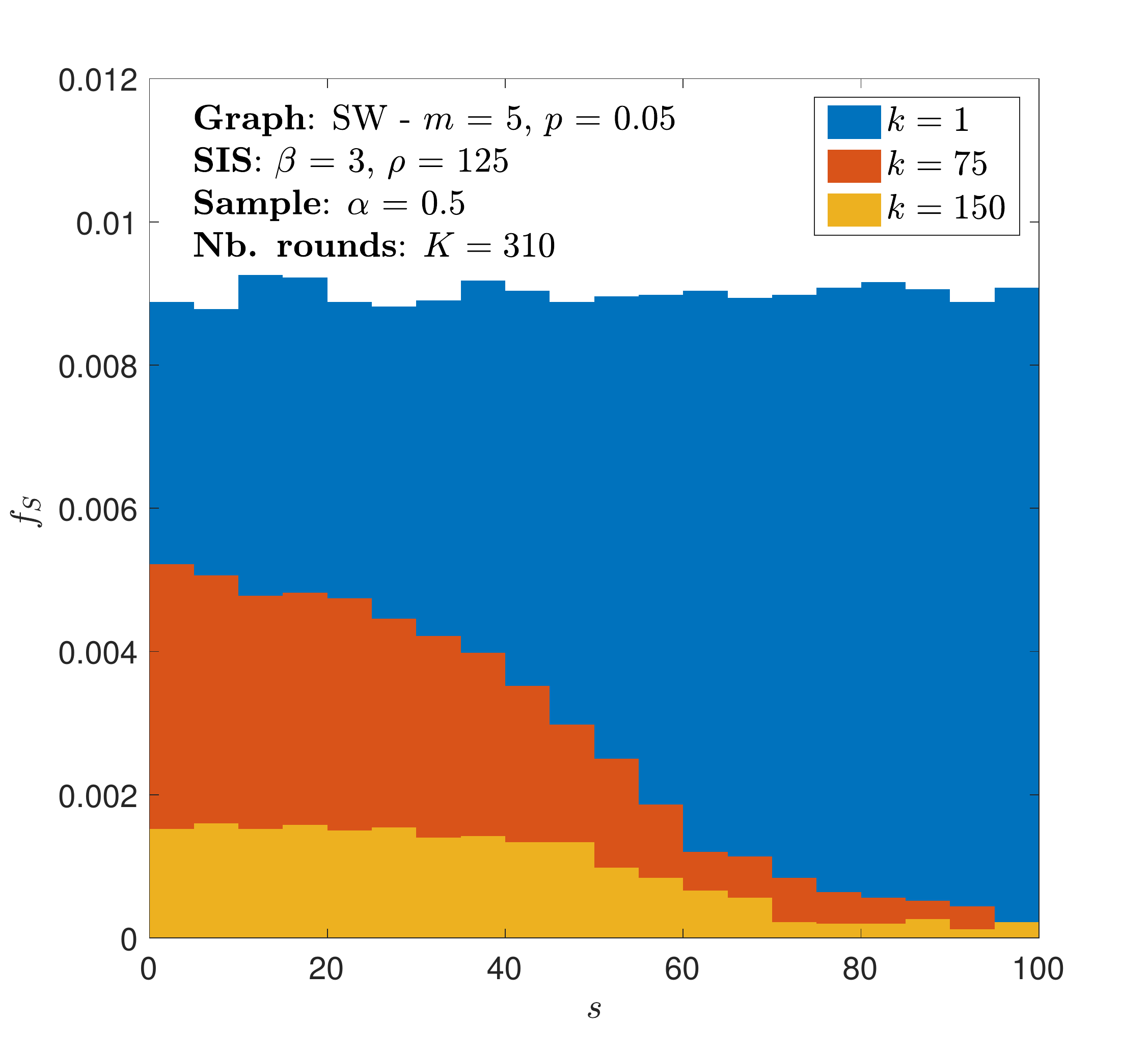}}\label{fig:---}}\\
\vspace{-1mm}
\vspace{-1mm}
\caption{The empirical \pdf $f_S$ of the node scores, at different different rounds $\t$ through the multi-round process, for a SW network. The results refer to the application of \RDRA or \SDRA (the \CCMstar) strategies with the MCM scoring function (initialization with the same infection level)).
}
\label{fig:score_distrib_mcm}
\vspace{-1mm}
\end{figure}

%---------------------------------------------------------
\inlinetitle{SW and SF networks}{.}
%---------------------------------------------------------
In \Fig{fig:perc_inf_mcm}, the simulations of \Sec{sec:online_comp} are repeated, changing only the scoring function which now is MCM (\RDRA appears in both sides for reference). 
The two scoring functions, LRIE and MCM, give similar results on SW and on SF networks. The noticeable difference concerns the \SDRA strategy for which the sequential \CCMstar (red curve) is almost identical to the \CCM strategy with $c=\sqrt{n}-1$ (green curve). In \cite{Bearden06}, the latter strategy is shown to be optimal for node scores drawn from a uniform distribution (but unobserved by the \DM who only makes pairwise comparisons). In order to verify this assertion, we displayed in \Fig{fig:score_distrib_mcm} the score distribution $f_S$ of the nodes \wrt the different rounds $k$. This clearly seems to be more uniform-like than the Gaussian-like shape obtained with LRIE, which explains well why the red and green curves are very close. Another observation is the fact that nodes with the highest scores are treated before scores with lowest scores (see orange bar plot), although some of them are still infected (the tail of the orange bar plot) since re-infection might occur at the beginning of the priority-order, especially in a graph without community structure. The results on a SF network are very similar to \Fig{fig:pa_perc_inf_left} and \Fig{fig:pa_perc_inf_right} and are therefore omitted due to space constraints. 

\begin{figure}[t]
\centering
\vspace{-2mm}
\hspace{-3mm}
\subfigure[User-user friendships dataset]{
\clipbox{12pt 0pt 0pt 0pt}
{\includegraphics[width = 0.485\linewidth, viewport=10 5 560 560, clip]{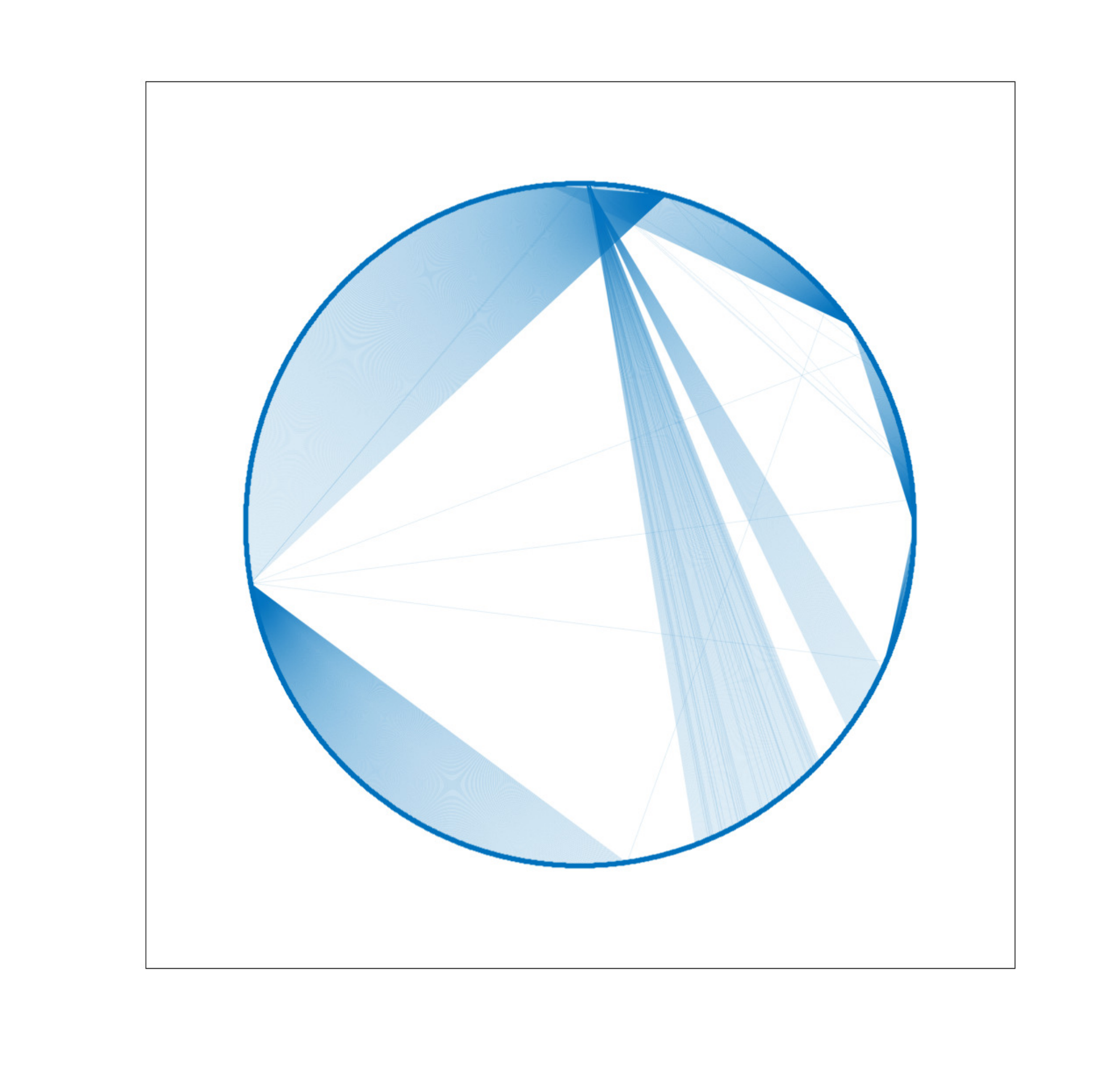}} }
\hspace{-2mm}
\subfigure[RDRA and SDRA performance]{
\clipbox{1.5pt -1pt 2pt 2pt}
{\includegraphics[width = 0.51\linewidth, viewport=0 142 580 690, clip]{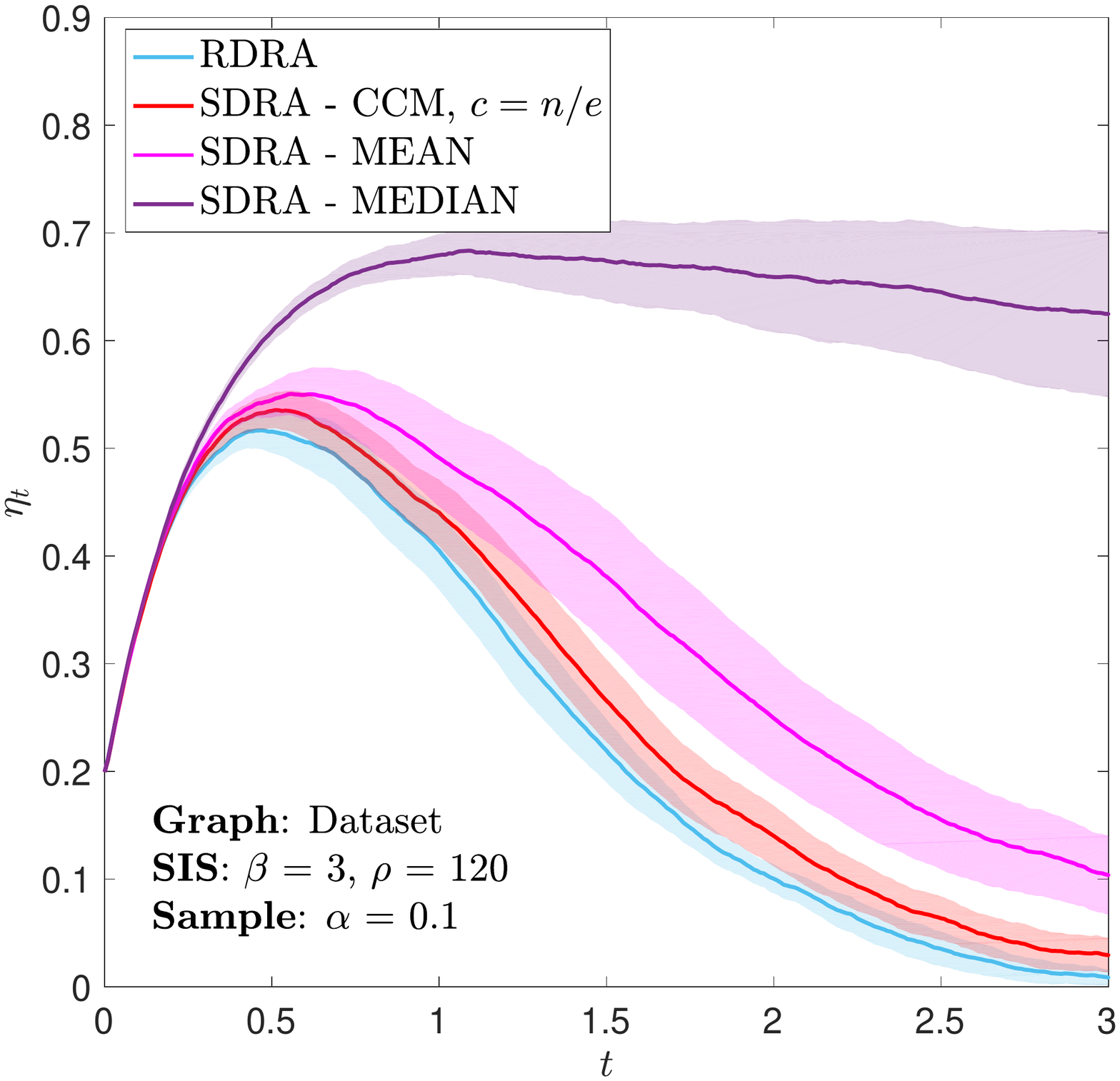}}}
\vspace{-2mm}
\caption{The average percentage of infected nodes \wrt time $t$, $\eta_t$ (right), on a real data network of $N=2888$ nodes from Facebook user-user friendships (left) using the MCM scoring function. The light blue curve displays the associated non-sequential \RDRA strategy.
The number of resources is $b=16$.} 
\label{fig:realdata}
\vspace{-1mm}
\end{figure}

%----------------------------------------------------------
\inlinetitle{Real networks}{.}
%----------------------------------------------------------
In the last simulations, in \Fig{fig:realdata}, we use a real network that contains Facebook user-user friendships\footnote{Available at: http://konect.uni-koblenz.de/networks/}, which is larger than the synthetic datasets used. A node is a user and an edge indicates a friendship between two users. Note that from \cite{Fekom19}, using \CCM with $c = \lfloor n/e\rfloor$ is a decent alternative to \CCMstar when few or no node recovered; it is therefore used in this simulation (red curve). 
We deduce, from the positions of nodes on the circle of the layout, that any node is easily reachable through a small number of jumps from another node, which is characteristic of a SW network. For the simulations, we use the MCM scoring function, slightly better than the LRIE (see Appendix).
Despite an initial number infected nodes of $20\%$, with only $b=16$ resources for $N = 2888$ nodes the epidemic exploses when the allocation strategy is not proper, \eg when using the \MEDIAN strategy (purple curve). 

%--------------------------------------------------------------------------------
\subsection{Offline vs. online} \label{sec:sim_offlineonline}
\begin{figure}[t]
\vspace{-3mm}
\centering
\hspace{-3mm}
\subfigure[$b/N = 0.05$]{
\clipbox{3pt 0pt 0pt 2.5pt}
{\includegraphics[width = 0.52\linewidth, viewport=2 8 555 570, clip]{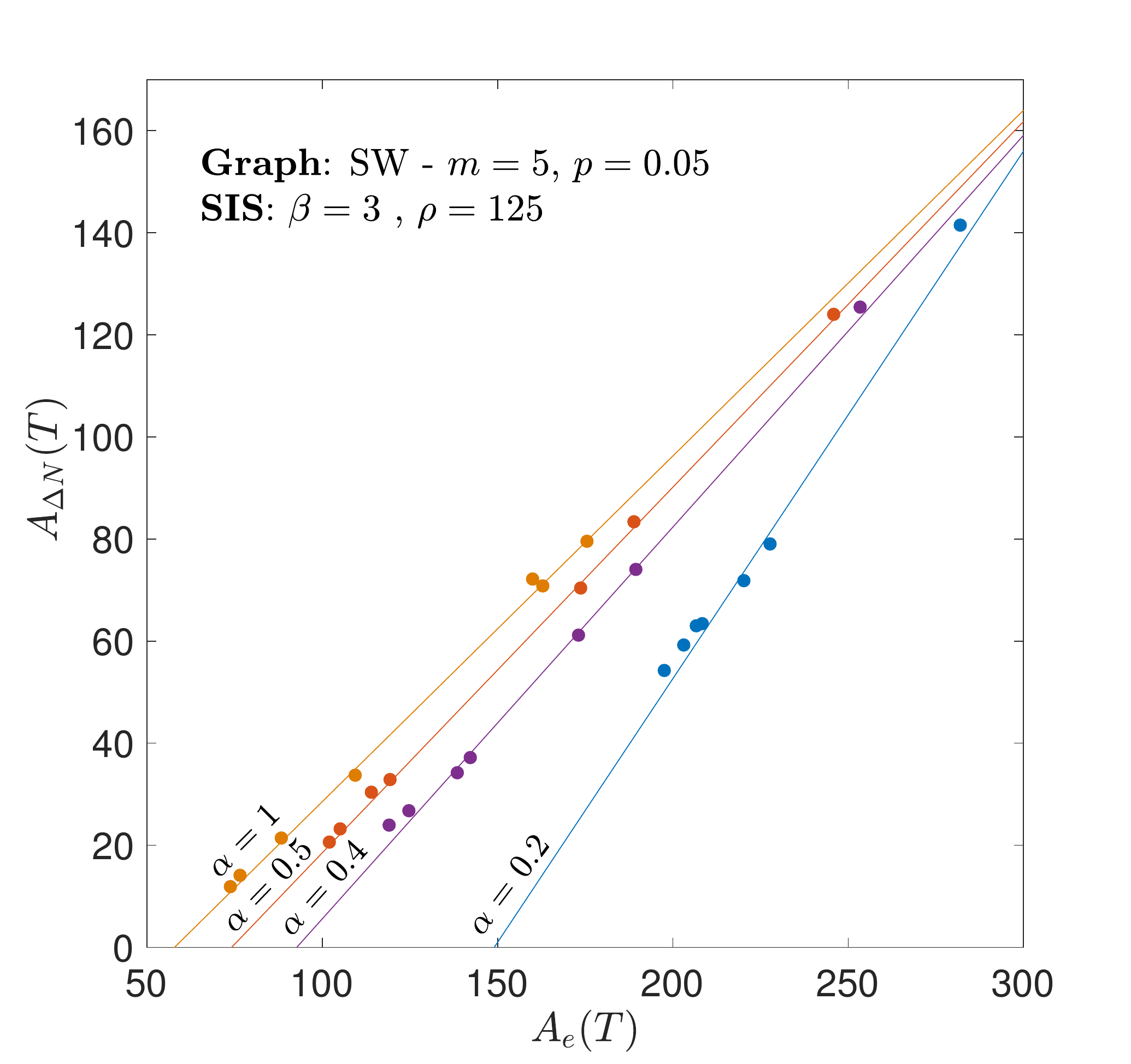}}\label{fig:linear_reg_b5}}
\hspace{-2mm}
\subfigure[$b/N = 0.06$]{ 
\clipbox{10pt 0pt 0pt 2.5pt}
{\includegraphics[width = 0.52\linewidth, viewport=2 8 555 570, clip]{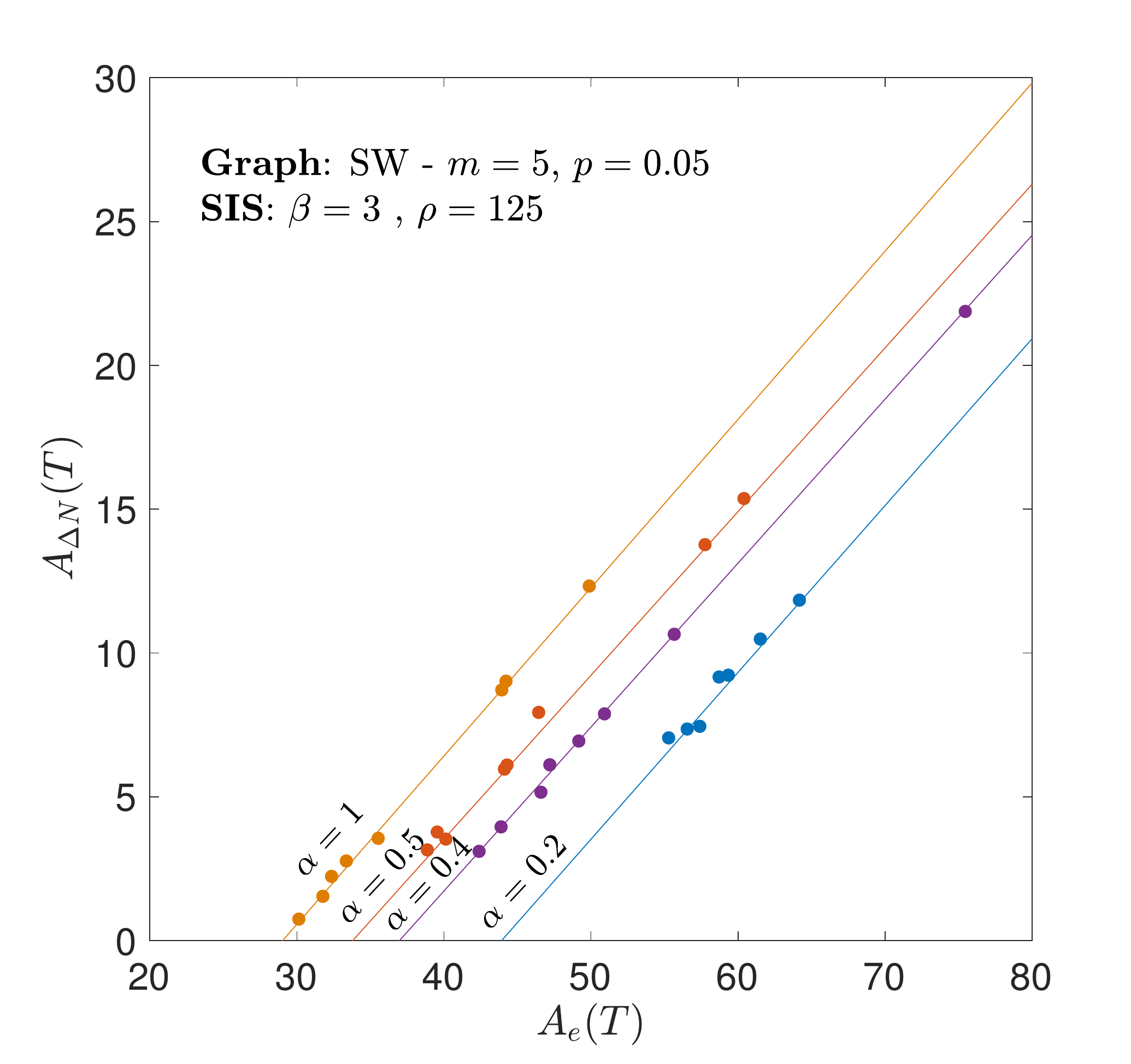}}\label{fig:linear_reg_b6}}\\ 
\vspace{-2mm}
\caption{At fixed time horizon $T$, the plots show the correlation between the AUC of the expected difference in percentage of infected using an online or the corresponding offline strategy, $A_{\diff N}(T)$ ($y$-axis) and the AUC of the expected number of errors, $A_e(T)$ ($x$-axis).}
\label{fig:linear_reg}
\vspace{-1mm}
\end{figure}

%--------------------------------------------------------------------------------
In this section we investigate the linear regression hypothesis stipulated in \Eq{eq:linear_reg} of \Sec{sec:off} by simulating the epidemic spread for different sequential selection strategies. We then compare, for a fixed time horizon $T$, the AUC of the expected percentage of infected nodes, $A_{\diff N}(T) = \int_0^T  \frac{\diff \numinfected_t}{N} dt$, and the AUC of the expected number of errors, $A_{e}(T) =  \sum_{k=1}^K \frac{\error_k}{b}$, as displayed in \Fig{fig:linear_reg}. The lines represent four different sampling sizes $\alpha = \{1, 0.5, 0.4, 0.2\}$ from left to right. 

\inlinetitleemph{Example}{ -- }
Imagine a scenario with $b=5$ treatments, $N=100$ nodes, and that the administrator examines sequentially half of them, \ie $\alpha=0.5$ (see \Fig{fig:linear_reg_b5}). As expected, the result of each strategy, \ie a 2-d point, lies on a line with slope coefficient $\cone=0.714$ and intercept $\ctwo=-52.14$; therefore we get $A_{\diff N}(T) \le 0.714 \cdot M_K-52.14$, which empirically verifies \Assumption{assum:linear_reg} (see \Tab{tab:linear_reg} in the Appendix for more examples).
After only 5-6 rounds, the \CCMstar algorithm makes no more than $1.5$ errors on average for a fixed number of candidates $n_{\max} = \alpha N = 50$ and $b=5$ resources, which gives $M_K\le \frac{1.5}{5} K = 0.3\cdot301 = 90.3$; hence we get $A_{\diff N}(T) \le 0.714\cdot90.3 - 52.14 = 12.33$ that should be compared with the maximum of AUC over $301$ rounds, that is $A_{\diff N}^{\text{MAX}}(T)=301$. Therefore, the error ratio is less than $\frac{A_{\diff N}(T)}{A_{\diff N}^{\text{MAX}}(T)} = 4.1 \%$ of the worse case scenario.   

%--------------------------------------------------------------------------------
\section{Conclusion and discussion}\label{sec:conclusion}
%--------------------------------------------------------------------------------

This study aimed towards bringing \DRAfull (\DRA) strategies closer to meeting real-life constraints.
We revisited their strong assumption that the \DM has full information and access to all network nodes, at any moment decisions takes place: anytime needed, she can instantaneously reallocate resources to any nodes indicated by a criticality scoring function.
We significantly relaxed this assumption by first introducing the \emph{\RDRAshort} model, where only a sample of nodes becomes accessible at each round of decisions.
Inspired by the way decisions are taken while care-seekers arrive at a healthcare unit, we next proposed the \emph{Sequential \DRA} model that limits further the control strategy so as to have only sequential access to the sample of nodes at each round. This setting offers a completely new perspective for dynamic control: the \DM examines the nodes one-by-one and decides \emph{immediately} and \emph{irrevocably} whether to reallocate resources or not. This online problem is put in relation with recent works in the literature where efficient algorithms have been presented for the selection of items from a sequence for which little or no information is available in advance. Special mention should be made to the \emph{\MSSPfull} (\MSSP) that has been found to be particularly fitting for handling this new setting.
Finally, according to our simulations on SIS epidemics, where we compared the performance of several variants of the above \DP control models, we conclude that the cutoff-based \CCMstar is a very promising approach for the setting of sequential \DP control.

\bibliographystyle{abbrv} 
\bibliography{SDRA_DP_v2_journal_arxiv}

\begin{thebibliography}{10}

\bibitem{Allen11}
D.~Allen, T.-C. Lu, D.~Huber, and H.~Moon.
\newblock Hierarchical random graphs for networks with weighted edges and
  multiple edge attributes.
\newblock In {\em \International \Conference on Data Mining}, 01 2011.

\bibitem{Allen08}
L.~Allen.
\newblock An introduction to stochastic epidemic models.
\newblock {\em Brauer F., van den Driessche P., Wu J. (eds) Mathematical
  Epidemiology. Lecture Notes in Mathematics, vol 1945. Springer, Berlin,
  Heidelberg.}, 2008.

\bibitem{Barabasi99}
A.-L. Barab{\'a}si and R.~Albert.
\newblock Emergence of scaling in random networks.
\newblock {\em Science}, 286(5439):509--512, 1999.

\bibitem{Bearden06}
J.~Bearden.
\newblock A new secretary problem with rank-based selection and cardinal
  payoffs.
\newblock In {\em \Journal of Mathematical Psychology}, volume~50, pages
  58--59, 2006.

\bibitem{Bekker17}
R.~Bekker, G.~Koole, and D.~Roubos.
\newblock Flexible bed allocations for hospital wards.
\newblock {\em Health Care Management Science}, 20(4):453--466, 2017.

\bibitem{Broder09}
A.~Z. Broder, A.~Kirsch, R.~Kumar, M.~Mitzenmacher, E.~Upfal, and
  S.~Vassilvitskii.
\newblock The hiring problem and lake {W}obegon strategies.
\newblock {\em SIAM \Journal on Computing}, 39:1223--1255, 2009.

\bibitem{Chen18}
D.~Chen, M.~Zheng, M.~Zhao, and Y.~Zhang.
\newblock A dynamic vaccination strategy to suppress the recurrent epidemic
  outbreaks.
\newblock {\em Chaos, Solitons \& Fractals}, 113:108--114, 2018.

\bibitem{Ezekiel20}
E.~Emanuel, G.~Persad, R.~Upshur, B.~Thome, M.~Parker, A.~Glickman, C.~Zhang,
  C.~Boyle, M.~Smith, and J.~Phillips.
\newblock Fair allocation of scarce medical resources in the time of covid-19.
\newblock {\em New England \Journal of Medicine}, 2020.

\bibitem{Fekom19bis}
M.~Fekom, N.~Vayatis, and A.~Kalogeratos.
\newblock Sequential dynamic resource allocation for epidemic control.
\newblock In {\em 2019 IEEE 58th \Conference on Decision and Control (CDC)},
  pages 6338--6343, 2019.

\bibitem{Fekom19}
M.~Fekom, N.~Vayatis, and A.~Kalogeratos.
\newblock The warm-starting sequential selection problem and its multi-round
  extension.
\newblock {\em arXiv preprint}, 2019.

\bibitem{Ferguson89}
T.~S. Ferguson.
\newblock Who solved the secretary problem?
\newblock {\em Statistical Science}, 1989.

\bibitem{Gnanlet09}
A.~Gnanlet and W.~G. Gilland.
\newblock Sequential and simultaneous decision making for optimizing health
  care resource flexibilities.
\newblock {\em Decision Sciences}, 40(2):295--326, 2009.

\bibitem{SISa_obesity}
A.~L. Hill, D.~G. Rand, M.~A. Nowak, and N.~A. Christakis.
\newblock Infectious disease modeling of social contagion in networks.
\newblock {\em PLoS Computational Biology}, 6(11):e1000968, 2010.

\bibitem{Jin2013}
F.~Jin, E.~Dougherty, P.~Saraf, Y.~Cao, and N.~Ramakrishnan.
\newblock Epidemiological modeling of news and rumors on twitter.
\newblock In {\em \Proceedings of the Workshop on Social Network Mining and
  Analysis}, 2013.

\bibitem{Kabene06}
S.~M. Kabene, C.~Orchard, J.~M. Howard, M.~A. Soriano, and R.~Leduc.
\newblock The importance of human resources management in health care: a global
  context.
\newblock {\em Human Resources for Health}, 4(1):20, 2006.

\bibitem{rumorSpreading2018}
A.~Kalogeratos, K.~Scaman, L.~Corinzia, and N.~Vayatis.
\newblock Information diffusion and rumor spreading.
\newblock In P.~M. Djuric and C.~Richard, editors, {\em Cooperative and Graph
  Signal Processing – {P}rinciples and Applications}, chapter~24, pages
  651--678. Elsevier, 2018.

\bibitem{Liu13}
M.~Liu and J.~Liang.
\newblock Dynamic optimization model for allocating medical resources in
  epidemic controlling.
\newblock {\em \Journal of Industrial Engineering and Management}, 6(1):73--88,
  2013.

\bibitem{Lloyd04}
A.~Lloyd.
\newblock Estimating variability in models for recurrent epidemics: Assessing
  the use of moment closure techniques.
\newblock {\em Theoretical population biology}, 65:49--65, 03 2004.

\bibitem{Lorch18}
L.~Lorch, A.~De, S.~Bhatt, W.~Trouleau, U.~Upadhyay, and M.~Gomez-Rodriguez.
\newblock Stochastic optimal control of epidemic processes in networks.
\newblock {\em Machine Learning for Health Workshop at NeurIPS}, 2018.

\bibitem{Mieghem09}
P.~V. Mieghem, J.~Omic, and R.~Kooij.
\newblock Virus spread in networks.
\newblock {\em IEEE/ACM \Transactions on Networking}, 17(1):1--14, 2009.

\bibitem{Scaman15}
K.~Scaman, A.~Kalogeratos, and N.~Vayatis.
\newblock A greedy approach for dynamic control of diffusion processes in
  networks.
\newblock In {\em IEEE \International \Conference on Tools with Artificial
  Intelligence}, pages 652--659, 2015.

\bibitem{Scaman16}
K.~Scaman, A.~Kalogeratos, and N.~Vayatis.
\newblock Suppressing epidemics in networks using priority planning.
\newblock {\em IEEE \Transactions on Network Science and Engineering},
  3(4):271--285, 2016.

\bibitem{Schneider11}
C.~M. Schneider, T.~Mihaljev, S.~Havlin, and H.~J. Herrmann.
\newblock Suppressing epidemics with a limited amount of immunization units.
\newblock {\em Physical Review E}, 84(6), 2011.

\bibitem{Tong12}
H.~Tong, B.~A. Prakash, T.~Eliassi-Rad, M.~Faloutsos, and C.~Faloutsos.
\newblock Gelling, and melting, large graphs by edge manipulation.
\newblock {\em Proceedings of the ACM CIKM}, pages 245--254, 2012.

\bibitem{Watts98}
D.~J. Watts and S.~H. Strogatz.
\newblock Emergence of scaling in random networks.
\newblock {\em Nature}, 393:440 EP, 1998.

\end{thebibliography}
\begin{table*}[!ht]
\centering 
\vspace{-1em}
\footnotesize
\resizebox{\textwidth}{!}{
\begin{tabular}{ l | l }
  \toprule
	\textbf{Symbol} & \textbf{Description}\\
	\midrule
	$\ind\{\mbox{condition}\}$ & indicator function\\
	$\one$ & vector with all values equal to one\\
	\midrule	
	$\mathcal{G}, \nodes, N, \mathcal{E}, E$ & network $\mathcal{G}= \{\nodes,\mathcal{E}\}$ of $N = |\nodes|$ nodes and $E = |\mathcal{E}|$ edges, where $\nodes$, $\mathcal{E}$ are the sets of nodes and edges\\
	$\adjbold \in \real^{N\times N}$							& network's adjacency matrix \st $\adj_{ij} \neq 0$ if node $i$ is linked with an edge to node $j$\\
	$t \in \real_+$ 		&  time\\
        $\Xbold{t} \in \spaceInfection^N$		& infection state vector \st $\X{i,t}{} = 1$ if node $i$ is infected at time $t$ \\
         $\mathbf{\bar{X}} \in \spaceInfection^N$		&  complementary of the infection state vector, \ie $\bar{X}_{i,t} = 1 - X_{i,t}$\\
	$\numinfected_t \in \mathbb{N}^*$		 & number of infected nodes at time $t$ \\
	\midrule	
	$b \in \mathbb{N}^*$ &   budget of control actions, here treatments allocated to nodes\\
	${\Abold}_t \in \{0,1\}^N$ &  resource vector at time $t$ \st $R_{i,t} = 1$ if a treatment is allocated to node $i$ at time $t$ subject to $||{\Abold}_t || = b$\\
	$\beta, \delta, \rho$ &  parameters of the \DP\\
	$s : \nodes \rightarrow \real$		 &  scoring function \st $S_{i,t} \in \real$ is the score of node $i$ at time $t$ \\
	$T \in \real_+$ & time-horizon\\
	$A_N(T) \in \real_+$ & Area Under the Curve of the percentage of infected nodes from $t=0$ to $t=T$\\
        $t_{\text{exct.}}$		& extinction time \ie smallest DP time for which $X(t_{exct})=\zero$ \\
   \midrule	
   $\Cobold{t}  \subset \nodes$ & the treated nodes at time $t$, called \preselection\\
   $k \le K$ & round index \st $t_k$ is the time at which the $k$-th round occurs, and $K$ is the total number of rounds\\
   $\policybold(\info,\access) \in \{0,1\}^n$ & control strategy for which the \DM has information on the nodes of the set $\info$, and access to the nodes of the set $\access$\\
	$n \le N, \Cbold{} \subset \nodes$ 		& node sample of size $n$ at time $t$ \st $\cone$ is the first incoming \candidate and $C_n$ the last of a round\\
	$\Lambda(c; n,x) \in [0,1]$ 	& probability  of observing \sample $c$ of size $n$ for a given infection state $x$ \\
	\midrule	
      $\cost \in \real_+$ & cost function to minimize\\
      $\Abold^\text{off} \in \{0,1\}^N$ & allocation vector that minimizes the cost for full access and info on the sample \ie $\info=\access=\Cbold{}$, subject to $||\Abold^\text{off}|| = b$\\
      $\Sobold, \mathbf{S^C} \subset \Sbold=s(\nodes)$  &  scores of the treated nodes, and of the candidate nodes respectively\\
    $\diff {\numinfected_t} \in \real_+$ &  difference between the expected number of infected nodes using an online strategy and the corresponding offline strategy\\
    $\error\in \real$ & online error, \ie the expected sum of half false negatives (FN) and false positives (FP) compared to an offline strategy\\
    $t_{\text{exct.}}$ & first time at which every nodes are healthy\\
    $\alpha \in [0,1]$  & fraction of the infected nodes that compose the sample\\
        $q \in ]0,1[$		& quality of the \preselection compared to the \candidates using the \CCM strategy\\
        $c \in \mathbb{N}$		& integer that specifies when to stop rejecting \candidates called cutoff using the \CCM strategy\\
         $\sigma_{N} : \real \times \real^N \rightarrow \{1,...,N\}$  &  ranking function, \st $\sigma_N(s, \mathbf{\Sigma})$ is the rank associated to score $s$ when compared to the scores in $S$\\
        $\cone, \ctwo$ 		& regression constants \st $\cone \geq 0 \geq \ctwo$\\
         $\ell$, $CUT(\ell)$ 			&  priority-order $\ell$, \st $\ell(i)$ is the order of node $i$ in the plan and max-cut of a given priority-order\\ 
          $f_S$ & distribution of the network nodes scores\\
	\bottomrule
\end{tabular}
}
\caption{Index of main notations.}
\label{tab:notations} 
\end{table*}

\newpage

%===============================================
\section*{Appendix}
%===============================================

%---------------------------------------------------------------------------------
\subsection{The dynamics of the SIS epidemic model}
%---------------------------------------------------------------------------------
\noindent As presented in the main text, the random variable of interest in the stochastic SIS epidemic model is the number of infected nodes at time $t$, $\numinfected_t$, more precisely its expectation $\Exp{\numinfected_t}$ for which we attempt to derive a closed-form equation.
Let us start by considering the following assumptions. 1)~The graph is a random \Erdos (ER) where every node has approximately the same degree, \ie $\bar{k} \approx k_i :=  \sum_{j=1}^N \adj_{ij},\,\, \forall i \in \nodes$.  2)~The allocation of resources is coarse-grained, \ie we apply the \RAND strategy for which the $b$ nodes to receive a treatment are chosen at random among the population of infected nodes. 3)~When $b>\numinfected_t$, the few remaining infected nodes can `accumulate' more than one resource to increase their recovery rate until $\numinfected_t = 0$. Note that those assumptions are usually not verified in real cases, but are taken as a simple starting point for the analysis. 
By respecting the above constraints, we multiply \Eq{eq:kolmo} by $\ni$, sum over $n$, and obtain:
\begin{align*}
 \frac{d\Exp{\numinfected_t}}{d t} &= \beta \textstyle\Exp{\sum_i  \textstyle \sum_j \adj_{ij} X_{j,t}\bar{X}_{i,t}}- \textstyle\Exp{ \sum_i(\delta + \rho R_{i,t})X_{i,t}}
\\
\frac{d\Exp{\numinfected_t}}{d t} &= \beta \,\,\Exp{\textstyle \sum_i \bar{k} \frac{\numinfected_t}{N} \bar{X}_{i,t} } - \delta \Exp{\numinfected_t} - \rho b\\
\frac{d\Exp{\numinfected_t}}{d t} &= \frac{\beta\bar{k}}{N}  \,\,\Exp{ \numinfected_t(N - N^ I_t)} - \delta \Exp{\numinfected_t} - \rho b,
\end{align*}
that finally leads to the evolution of the first-order moment:
\begin{equation}
\frac{d\Exp{\numinfected_t}}{d t} = (\beta \bar{k}-\delta)\,\,\Exp{\numinfected_t} - \frac{\beta\bar{k}}{N}\,\,\Exp{(\numinfected_t)^2}  - \rho b.
\label{eq:first_order}
\end{equation}
We now multiply \Eq{eq:kolmo} by $n^2$, sum over $n$, and similarly obtain the dynamic equation for the second-order moment:
\begin{align}
\begin{split}
\frac{d\Exp{(\numinfected_t)^2}}{d t} &= (\beta \bar{k}+\delta-2\rho b)\,\Exp{\numinfected_t} -2 \frac{\beta\bar{k}}{N}\,\,\Exp{(\numinfected_t)^3}\\
 &+ \left( 2(\beta \bar{k}- \delta)-\frac{\beta \bar{k}}{N}\right)\Exp{(\numinfected_t)^2} + \rho b. 
\end{split}
\label{eq:second_order}
\end{align}
It is easy to check that every moment equation depends on a higher order moment, and thus the closed-form equation for the evolution of the expected number of infected nodes cannot be computed, even in the simple coarse-grained allocation case.
In order to get some results, a first option is to use a moment closure technique to approximate the highest-order term, as in \cite{Lloyd04}.
The two most common approximations are the following:
\begin{align}
 &\text{Normal: } \Exp{(\numinfected_t)^3} = 3\Exp{(\numinfected_t)^2}\Exp{(\numinfected_t)} - 2(\Exp{(\numinfected_t)})^3\\
  &\text{Lognormal: }\Exp{(\numinfected_t)^3} = \left(\frac{\Exp{(\numinfected_t)^2}}{\Exp{(\numinfected_t)}}\right)^3\!.\label{eq:lognormal}
 \end{align}
A second option is to bound the expected number of infected nodes by using the fact that $\Exp{(\numinfected_t)^2} \ge {\Exp{\numinfected_t}}^2$, and thus that the mean of the stochastic SIS epidemic model is less than that of the deterministic solution.
\begin{figure}[t]
\centering
\hspace{-3mm}
\subfigure[SW network]{
\clipbox{0pt 0pt 0pt 2pt}
{\includegraphics[width = 0.51\linewidth, viewport=0 130 580 690, clip]{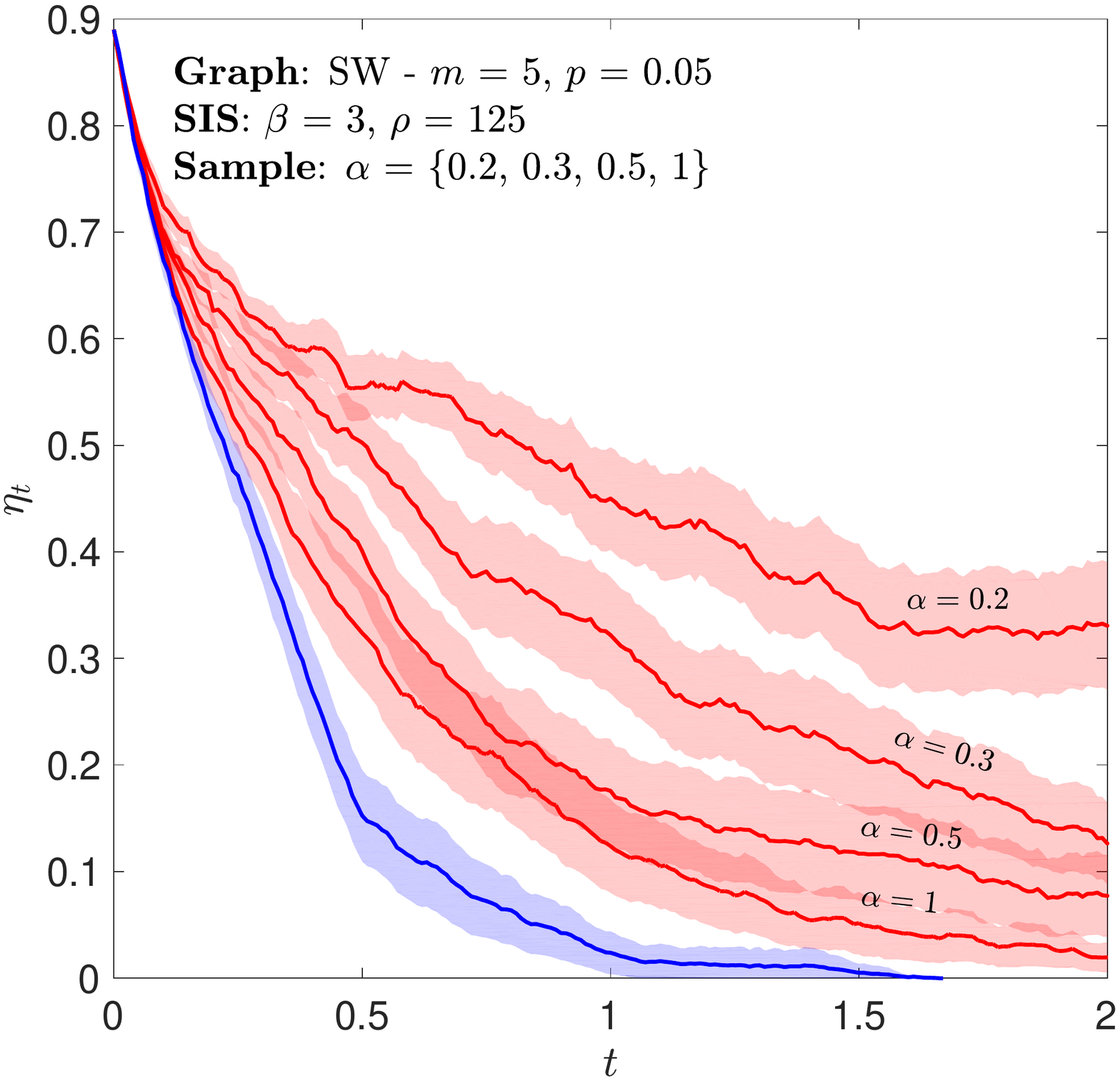}}\label{fig:sample_size_SW}}
\hspace{-3mm}
\subfigure[SF network]{ 
\clipbox{5pt 0pt 0pt 2pt}
{\includegraphics[width = 0.51\linewidth, viewport=0 130 580 690, clip]{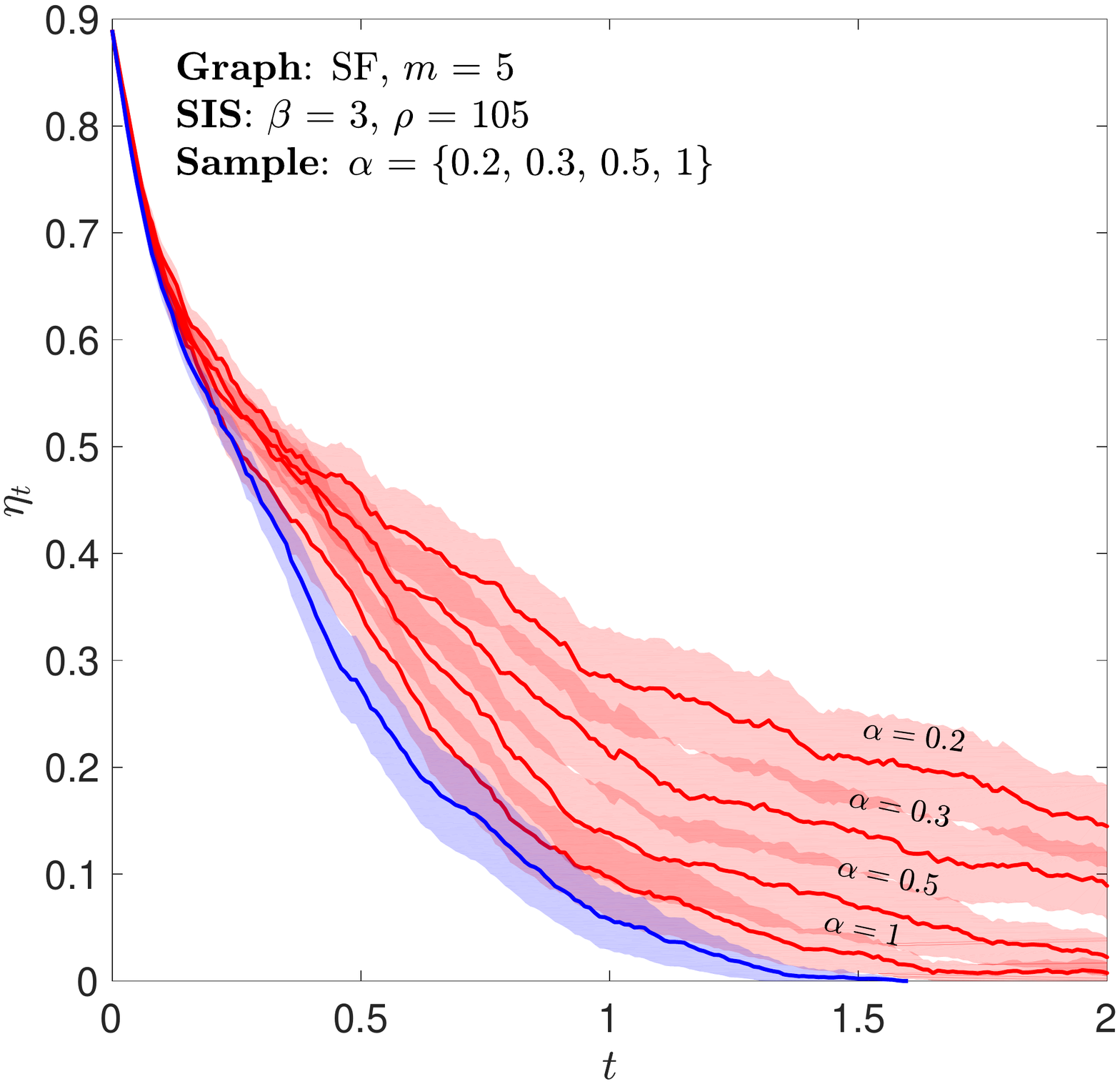}}\label{fig:sample_size_PA}} 
\vspace{-2.5mm}
\caption{The average percentage of network infection though time using the sequential \CCMstar strategy (red lines), for various fixed sampling sizes. The blue curves display the associated non-sequential \RDRA strategy with full access to nodes at each round (\ie $\alpha=1$).}
\label{fig:sample_size}
\end{figure}
\begin{figure}[ht]
\vspace{-5.2mm}
\centering
\hspace{-5mm}
\subfigure[$\bar{k} = 2$]{
\clipbox{1pt 0pt 2pt 3.5pt}{
\includegraphics[width=0.518\linewidth, viewport=20 8 555 530,clip]{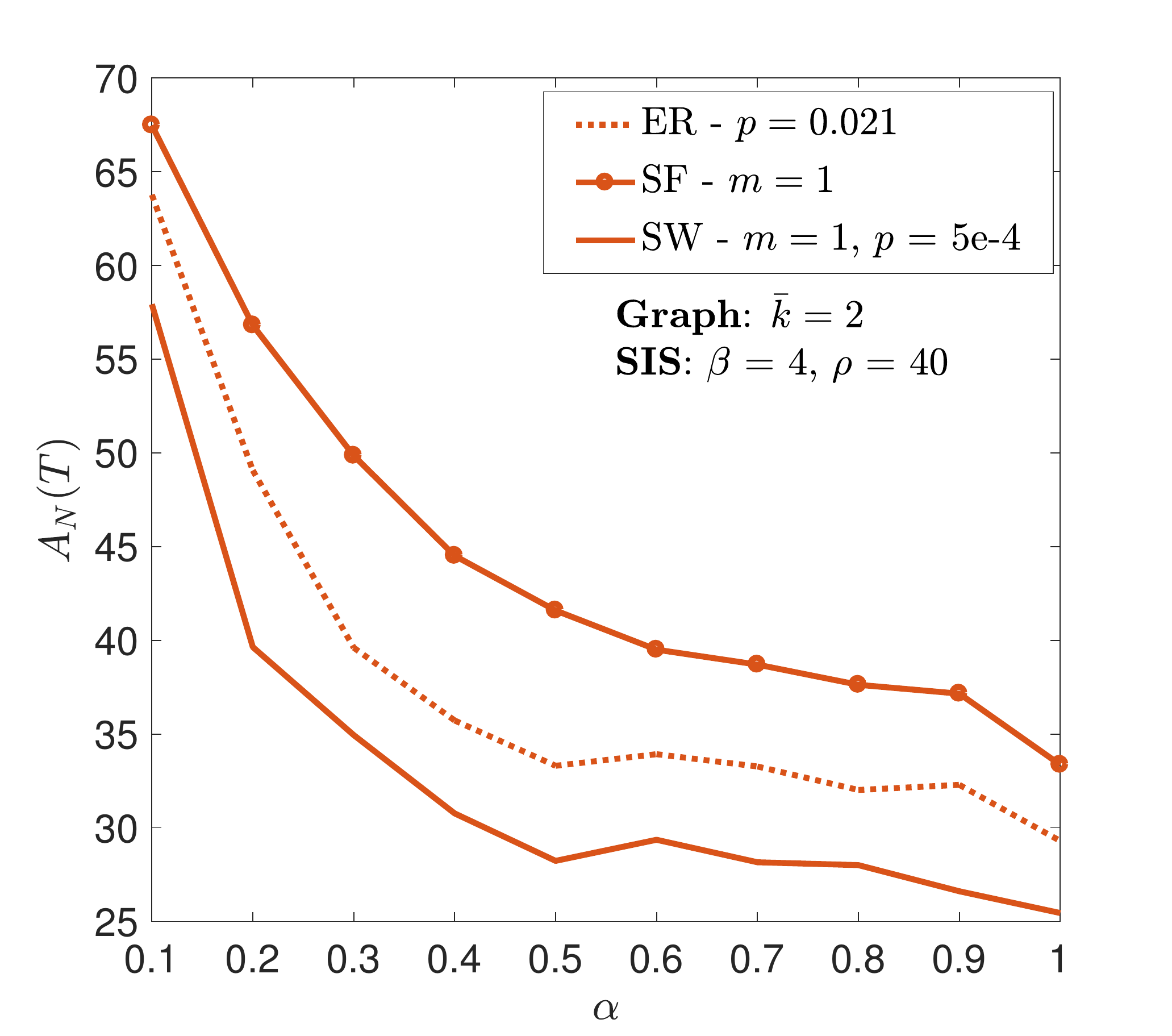} }}
\hspace{-4mm}
\subfigure[$\bar{k} = 10$]{
\clipbox{8pt 0pt 4pt 3.5pt}{
\includegraphics[width=0.518\linewidth ,viewport=20 8 555 530,clip]{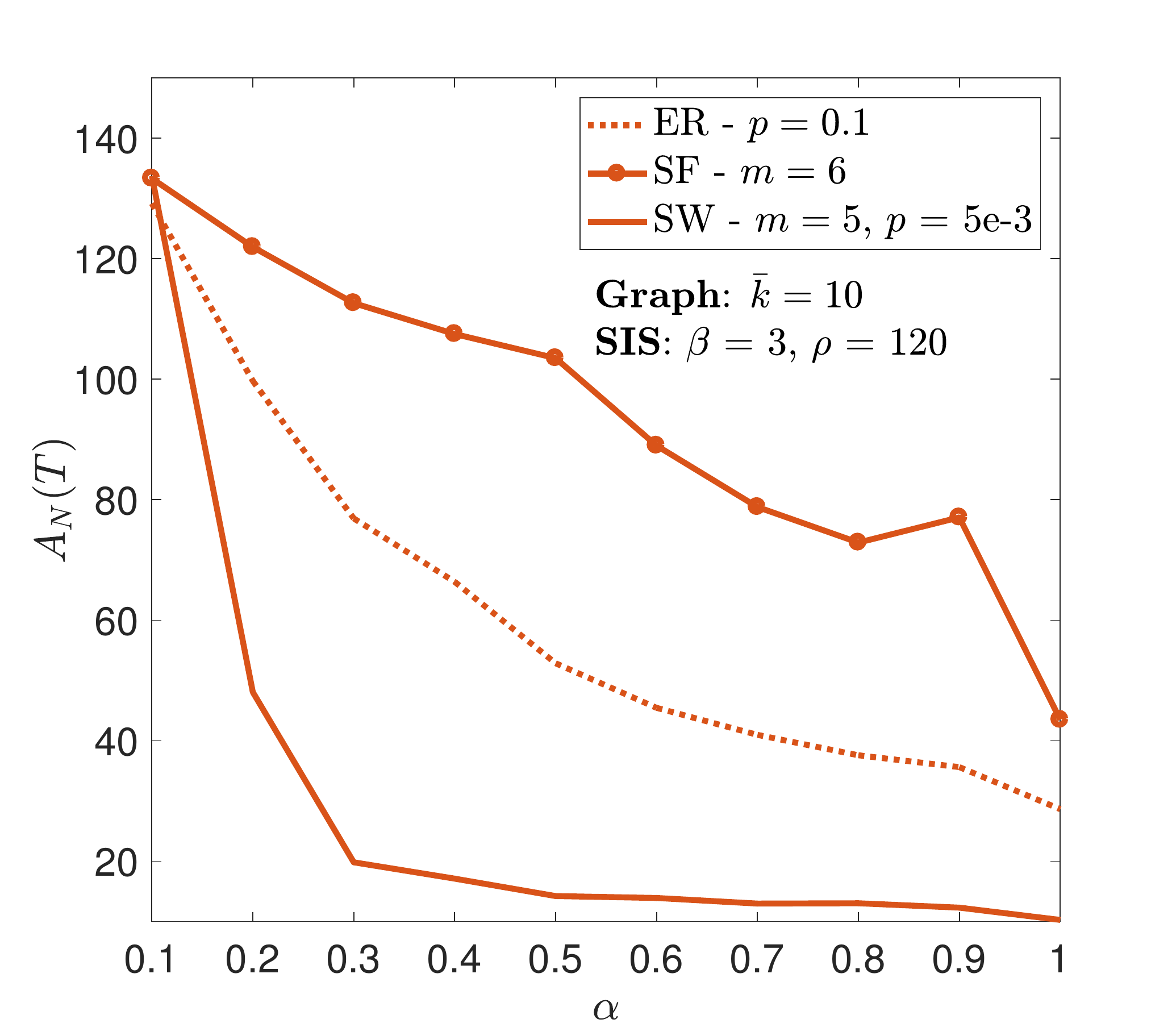} }}
\vspace{-2mm}
\caption{AUC \wrt $\alpha$, \ie the ratio of infected nodes that get accessible at each round, with $b=5$ resources and $N=100$ nodes. Dark, medium and light orange respectively stand for an \Erdos (ER), a scale-free (SF), and a small-world (SW) network, each of them having an average degree of $\bar{k}$.}
\label{fig:auc}
\end{figure}

%-------------------------------------------------------------------------------
\subsection{Sampling size}
%--------------------------------------------------------------------------------
As described, the sampling is performed on the infected nodes and so far we used an arbitrary fixed ratio $\alpha$. To analyze the impact of the sampling size on the efficiency of the \CCMstar strategy, we plot in \Fig{fig:sample_size} the average percentage of infected nodes \wrt time for various sampling ratios.
We observe that the SDRA is less sensitive to the sampling size on SF networks (right) than on SW networks (left). Furthermore, regardless the network structure, increasing the sampling size does not improve linearly the efficiency of the algorithm.
In \Fig{fig:auc} is displayed the AUC of \Eq{eq:auc} for two different number of average neighbors $\bar{k}=\{2,10\}$, and on each figure, for three different types of networks. Clearly, the difference between the network types is more evident when the edge density of the network increases (right-hand side), and the AUC is smaller for the SW type. This can be explained by the fact that in this type of graphs, increasing the edge density merely reinforces the connectivity within the hubs, \ie aggregate of nodes, and hence assigning resources to those critical nodes allows to reduce efficiently the spread.
Note that, in this work we consider a passive sampling that is purely random and does not follow any strategy, however we might envision an adversarial sampling made by a malicious agent that adapts to the \DM's strategy. In this case, the worst case scenario plays an important role in the strategy's performance, using the \CCM strategy for instance, it occurs when the $b$-best candidates are the first incoming, since they are very likely to be rejected by default.

%-------------------------------------------------------------------------------
\subsection{Simulations support} 
\begin{figure}[]
\centering
\hspace{-7mm}
\subfigure[\scriptsize Network]{
\clipbox{3pt 6pt 2pt 1pt}{
\includegraphics[width=0.52\linewidth, viewport=5 8 530 530,clip]{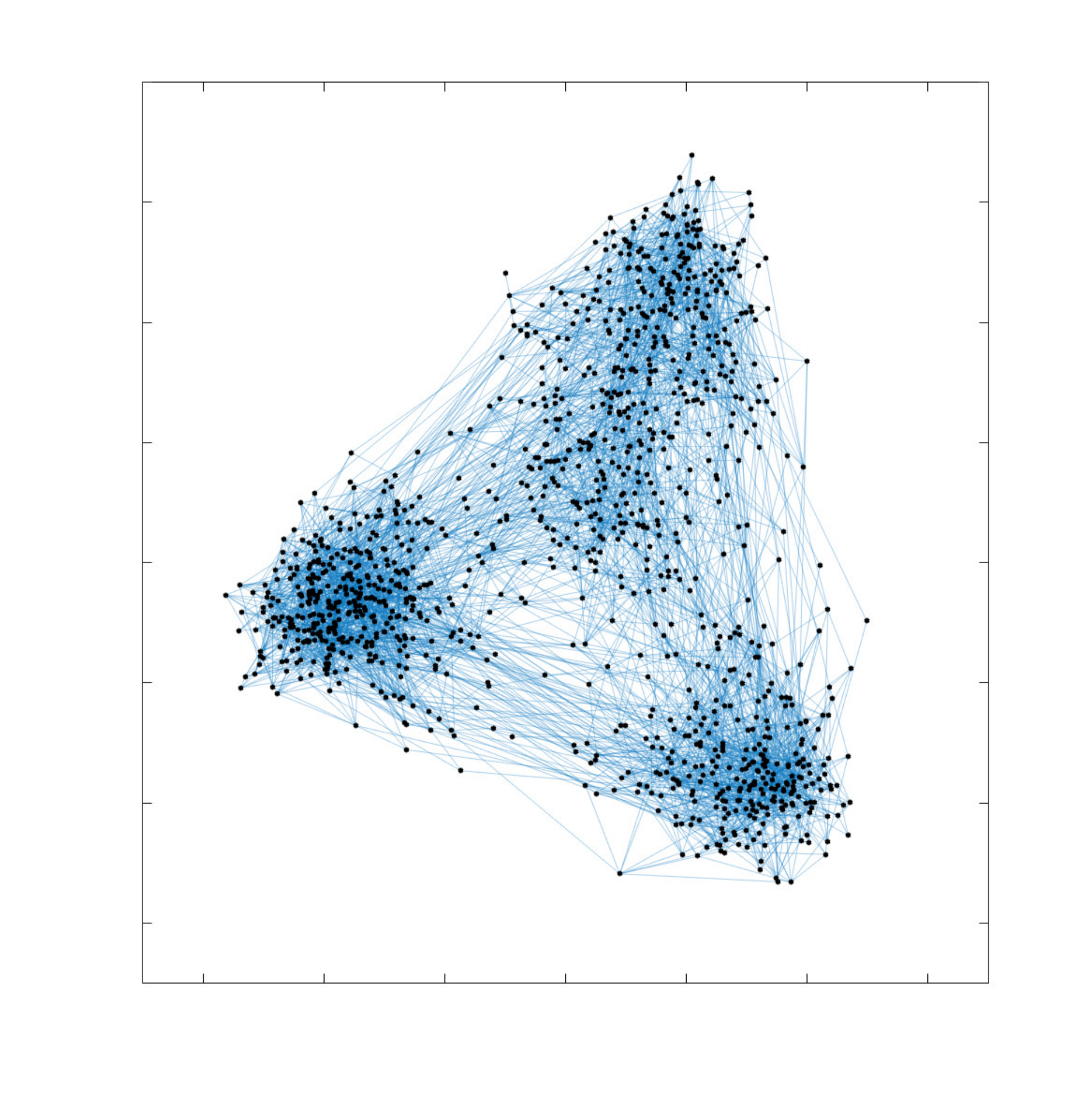} }}
\hspace{-4mm}
\subfigure[\scriptsize Block-diagonal adjacency matrix]{
\clipbox{7.5pt 6pt 4pt 3pt}{
\includegraphics[width=0.53\linewidth ,viewport=5 8 530 530,clip]{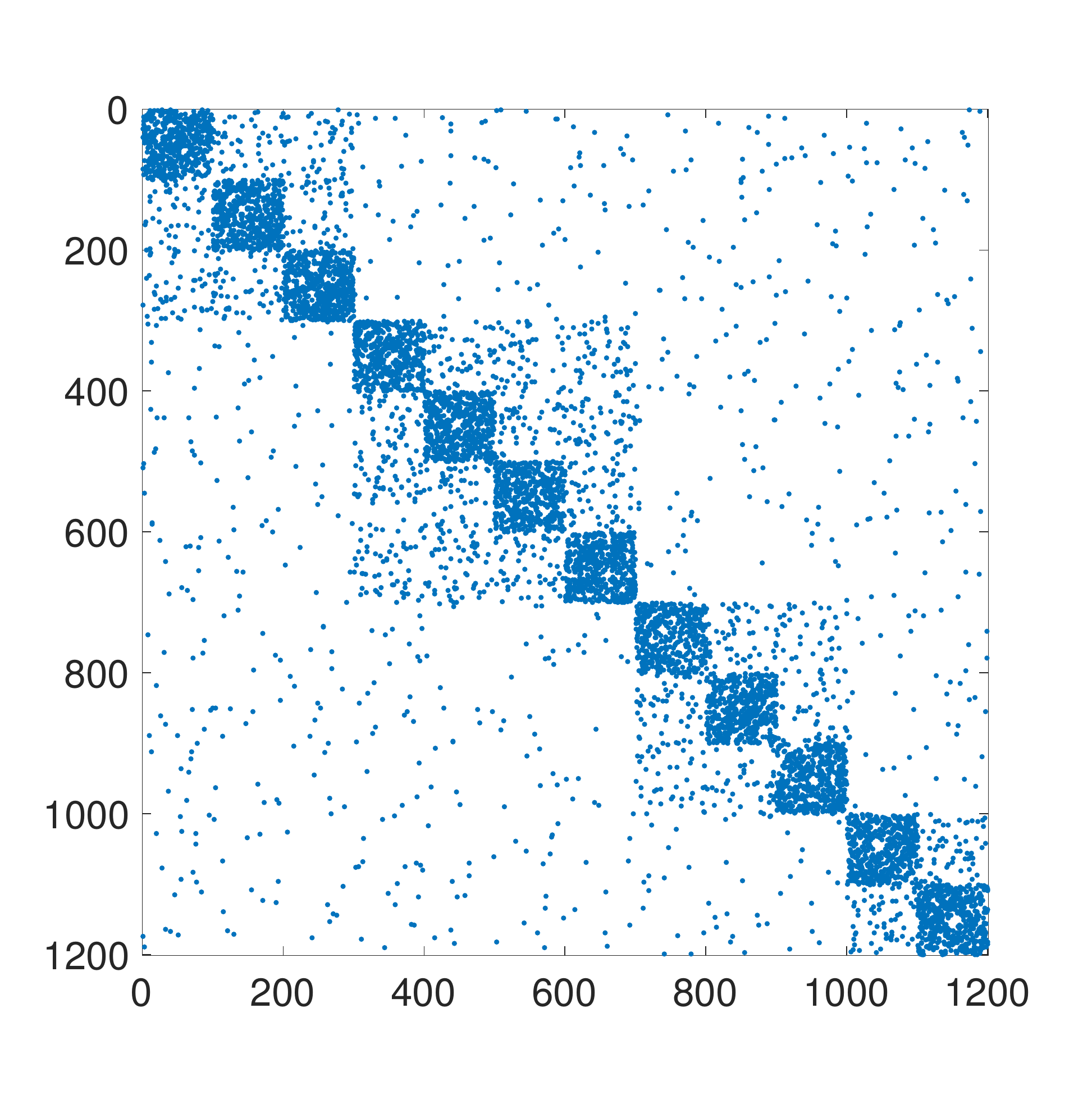} }}
\caption{Example of community-structured (CS) network with $N=1200$ nodes and $12$ groups.}
\label{fig:community}
\end{figure}

%--------------------------------------------------------------------------------
%--------------------------------------------------------------------------------
\inlinetitle{Community-structured network}{.} 
%--------------------------------------------------------------------------------
The community graph used for simulations in \Fig{fig:score_functions} is displayed in \Fig{fig:community}, showing a clear hierarchical structure with three levels of point density. 

\begin{figure}[]
\centering
\vspace{2.2mm}
\hspace{-4mm}
\subfigure[\scriptsize LRIE \vs MCM scoring function]{ 
\clipbox{0pt 0pt 2pt 2pt}
{\includegraphics[width = 0.51\linewidth, viewport=0 130 580 690,clip]{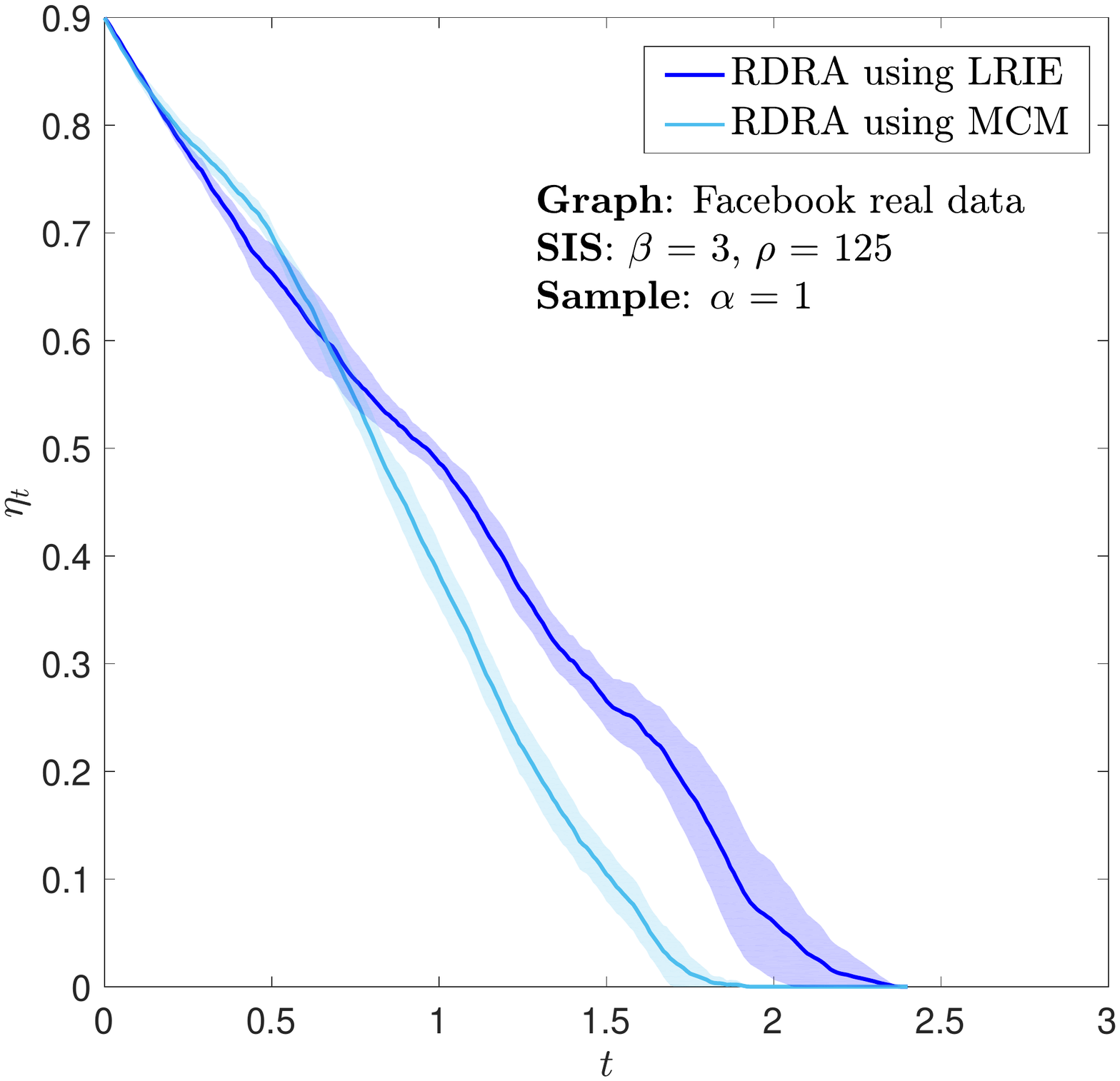}}}
\subfigure[\scriptsize LRIE scoring function]{ 
\clipbox{4.5pt 0pt 2pt 2pt}
{\includegraphics[width = 0.51\linewidth, viewport=0 130 580 690, clip]{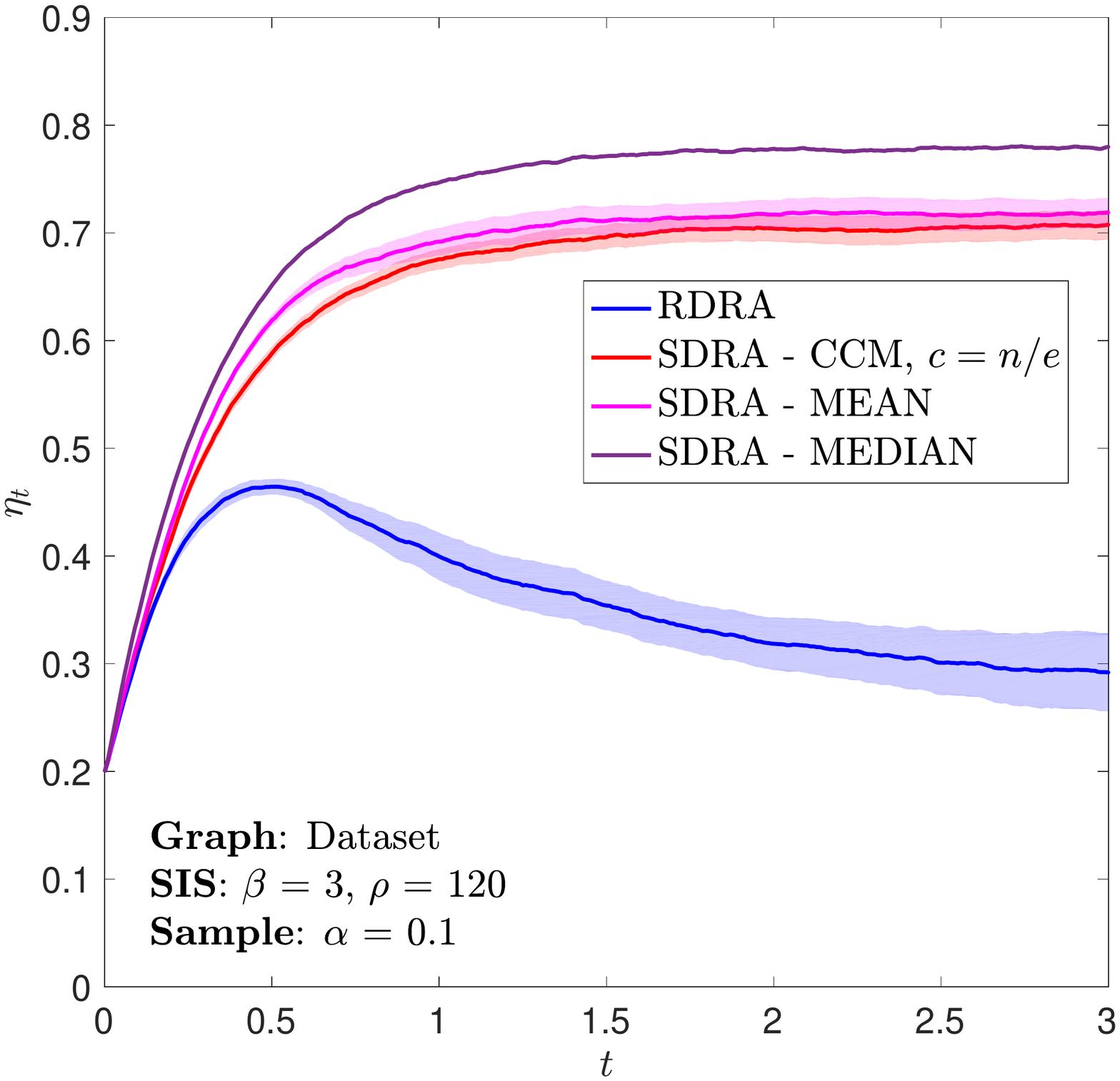}}}
\vspace{-2mm}
\caption{The average percentage of infected nodes \wrt time $t$, $\eta_t$, on a real data network of $N=2888$ nodes from Facebook user-user friendships. The number of resources is $b=16$.}
\label{fig:realdataLRIE}
\end{figure}

%--------------------------------------------------------------------------------
\inlinetitle{Real network}{.} 
%--------------------------------------------------------------------------------
The simulations of \Fig{fig:realdataLRIE} are similar to those of \Fig{fig:realdata} but focusing on the LRIE instead of the MCM scoring function. On the left, \RDRA strategies using both scoring functions are compared, while on the right a comparison of different sequential strategies using LRIE is presented. Contrary to \Fig{fig:realdata} with MCM, the sequential strategies here fail to reduce the epidemic spread when using the LRIE scores.
%-------------------------------------------------------------------------------
\inlinetitle{Regression parameters}{.} 
%--------------------------------------------------------------------------------
The values used for computing the example in \Sec{sec:sim_offlineonline} are shown in \Tab{tab:linear_reg}.
\begin{table}[]
\centering
\vspace{-5mm}
\footnotesize
\begin{tabular}{lcccc}
\hline
               & \multicolumn{2}{l}{$\ b/N = 0.05$} & \multicolumn{2}{l}{$\ b/N = 0.06$} \\
               \hline
               & $\cone$           & $\ctwo$          & $\cone$           & $\ctwo$          \\
               \hline
$\alpha = 1$   & 0.68            & -39         \ \ \    &      0.584           &    -18.7            \\
$\alpha = 0.5$ & 0.714           & -52.4     \ \ \  &    0.579             &       -19.2         \\
$\alpha = 0.2$ & 1.03            & -153    \ \ \       &       0.58          &  -  25.4           \\
\hline
\end{tabular}
\vspace{1mm}
\caption{\scriptsize Values of constants $\cone$ and $\ctwo$ for $\beta=3$, $\rho=125$ for a SW graph with $p=0.05$ and $m=5$.}
\label{tab:linear_reg} 
\vspace{-6mm}
\end{table}
\end{document}